\documentclass[conference]{IEEEtran}
\IEEEoverridecommandlockouts
\usepackage{cite}
\usepackage{amsmath,amssymb,amsfonts}
\usepackage{algorithmic}
\usepackage{graphicx}
\usepackage{textcomp}
\usepackage{xcolor}
\def\BibTeX{{\rm B\kern-.05em{\sc i\kern-.025em b}\kern-.08em
    T\kern-.1667em\lower.7ex\hbox{E}\kern-.125emX}}

\usepackage[normalem]{ulem}

\newtheorem{theorem}{Theorem}\newtheorem{corollary}{Corollary}[theorem]
\newtheorem{lemma}[theorem]{Lemma}

\usepackage{enumitem}
\usepackage{hyperref}

\usepackage[ruled,vlined,linesnumbered,norelsize]{algorithm2e}
\usepackage{algorithmic}

\usepackage{dblfloatfix}

\newcommand{\suppl}{Appendix}

\newcommand{\algname}{\textit{Leashed-SGD}}
\newcommand{\asyncsgd}{\emph{AsyncSGD}\/}
\newcommand{\hogwild}{\textsc{Hogwild!}\/}
\newcommand{\syncsgd}{\emph{SyncSGD}\/}
\newcommand{\paramcont}{\emph{ParameterVector}\/}
\newcommand{\casretryloop}{\emph{LAU-SPC}\/}

\begin{document}

\title{Consistent Lock-free Parallel Stochastic Gradient Descent for Fast and Stable Convergence
}

\author{\IEEEauthorblockN{Karl Bäckström, Ivan Walulya, Marina Papatriantafilou, Philippas Tsigas}
\IEEEauthorblockA{Dept. of Computer Science and Engineering, Chalmers University of Technology, Gothenburg, Sweden\\
\{bakarl, walulya, ptrianta, tsigas\}@chalmers.se
}
}

\maketitle
\begin{abstract}
\textit{Stochastic gradient descent} (SGD) is an essential element in Machine Learning (ML) algorithms.
Asynchronous parallel shared-memory SGD (\asyncsgd{}), including synchronization-free algorithms, e.g. \hogwild{}, have received interest in certain contexts, due to reduced overhead compared to synchronous parallelization.
Despite that they induce staleness and inconsistency, 
they have shown speedup for problems satisfying smooth, strongly convex targets, and gradient sparsity.
Recent works take important steps towards understanding the potential of parallel SGD for problems not conforming to these strong assumptions, in particular for \textit{deep learning} (DL).
There is however a gap in current literature in understanding when \asyncsgd{} algorithms are useful in practice, and in particular how mechanisms for synchronization and consistency play a role.

We contribute with answering questions in this gap
by studying a spectrum of parallel algorithmic implementations of \asyncsgd{},
aiming to understand how shared-data synchronization influences the convergence properties in fundamental DL applications. We focus on the impact of consistency-preserving non-blocking synchronization in SGD convergence, and in sensitivity to hyper-parameter tuning.
We propose \algname{}, an extensible algorithmic framework of consistency-preserving implementations of \asyncsgd{}, employing lock-free synchronization, effectively balancing throughput and latency. \algname{} features a natural contention-regulating mechanism, as well as dynamic memory management, allocating space only when needed.
We argue analytically about the dynamics of the algorithms, memory consumption, the threads' progress over time, and the expected contention. The analysis further shows the contention-regulating mechanism that \algname{} enables.

We provide a comprehensive empirical evaluation, validating the analytical claims,
benchmarking the proposed \algname{} framework, and comparing to baselines for two prominent \textit{deep learning} (DL) applications: \textit{multilayer perceptrons} (MLP) and \textit{convolutional neural networks} (CNN). We observe the crucial impact of \textit{contention}, \textit{staleness} and \textit{consistency} and show how, thanks to the aforementioned properties, \algname{} provides
significant improvements in stability as well as wall-clock time to convergence (from 20-80\% up to $4\times$ improvements)
compared to the standard lock-based \asyncsgd{} algorithm and \hogwild{}, while reducing the overall memory footprint.

\end{abstract}

\section{Introduction}

The interest in Machine Learning (ML) methods for data analytics has peaked in the last decade due to their tremendous impact across various applications.
Parallel algorithms for ML, utilizing modern computing infrastructure, have gained particular interest, showing high scalability potential, necessary in accommodating for significant growing data demands as well as data availability.
Parallelization schemes for Stochastic Gradient Descent (SGD) have been of particular interest, since SGD serves as a backbone in many widely used ML algorithms and has proven effective on convex problems (e.g. linear, logistic regression, SVM), as well as non-convex (e.g. matrix completion, deep learning).
\looseness=-1

The first-order iterative minimizer SGD follows the simple rule (\ref{eq:sgd_step}) of moving in the direction of the negative stochastic gradient $\widetilde{\nabla}f$ with a step size $\eta$, of a differentiable target function $f: \mathbb{R}^d \rightarrow \mathbb{R}$, quantifying the error of a ML model:
\begin{align}
    \theta_{t+1} = \theta_t - \eta \widetilde{\nabla} f(\theta_t) \label{eq:sgd_step}
\end{align}
where $\theta_t$ contains the \textit{learned} parameters of the model at iteration~$t$, typically encoding features of a given data-set.  Iterations, calculating over \emph{batches} of one or multiple data samples each, typically repeat until \emph{$\epsilon$-convergence}, i.e. reaching a sufficiently low error threshold $\epsilon$.
As in SGD each update relies on the outcome of the previous one, data parallelization is challenging. Still, several approaches have been proposed, distinguished into \textit{synchronous} and \textit{asynchronous} ones: \looseness=-1 \\
\textit{Synchronous SGD} (\syncsgd) is a lock-step parallelization scheme where the gradient computation is delegated to threads/nodes, then aggregated by averaging before taking a global step according to eq.~(\ref{eq:sgd_step})~\cite{zinkevich2010parallelized}.
In its original form, \syncsgd{} is statistically equivalent to sequential SGD with larger \textit{data-batch}\cite{gupta2016model}\cite{backstrom2019mindthestep}.
This method is well-understood and widely used, e.g. in 
\textit{federated learning}~\cite{mcmahan2017communication}.
However, its scalability suffers as every step is limited by the slowest contributing thread. In addition, higher parallelism implies an impact on the convergence, inherent to \textit{large-batch} training~\cite{keskar2016large}.
Semi-synchronous variants have shown improvements~\cite{lee2014model,li2020taming}, relaxing lock-step semantics and requiring only a subset of threads to synchronize, hence reducing waiting.
In a recent article~\cite{li2020taming} it was seen that requiring only a few, even just one, thread at synchronization, implies significant speedup due to less waiting and higher throughput, motivating further study of \textit{asynchronous} parallel SGD.\\
\textit{Asynchronous SGD} (\asyncsgd) on the other hand employs parallelism on SGD/algorithm level, allowing threads to execute (\ref{eq:sgd_step}) on a shared vector $\theta$ with less coordination, and has shown superior speedup compared to \syncsgd{} in several applications~\cite{recht2011hogwild,ma2018stochastic}. It was first introduced for distributed optimization with a parameter server sequentializing the updates. In this context it was proven that the algorithm converges for convex problems \cite{agarwal2011distributed} despite the presence of noise due to stale updates.
A relaxed variant, \hogwild{}~\cite{recht2011hogwild}, allowing completely uncoordinated component-wise reads and updates in $\theta$, showed substantial speedup, however only on smooth convex problems with sparse gradients.
This, besides \emph{staleness}, also introduces \textit{inconsistency} incurred by non-coordinated concurrent reads and writes on $\theta$, penalizing the statistical efficiency. Only if parallelization gains counterbalance the latter penalty, will there be an actual improvement in the wall-clock time for convergence.\looseness=-1

\smallskip 

\subsubsection*{Challenges}
There are substantial analytical results and empirical evidence that \asyncsgd{} \cite{agarwal2011distributed,chaturapruek2015asynchronous,duchi2015asynchronous,recht2011hogwild} provides speedup for problems satisfying varying assumptions on convexity, strong convexity, smoothness and sparsity, e.g. Logistic regression, Matrix completion, Graph cuts and SVM training.
Recently, a target of study is parallelism in SGD for wider class of more unstructured problems, not conforming to strict analytical assumptions, such as \textit{artificial neural network} (ANN) training, or \textit{deep learning} (DL) in general.
Recent works \cite{ben2019demystifying, dutta2018slow} explore aspects of data-parallelism in the context of distributed and parallel SGD for DL.
However, using abstraction libraries such as TensorFlow and Keras in Python implementations, with its inherent limitations in parallelism and performance, makes time measurements unreliable. As a consequence, the existing literature address the topic mostly from an analytical standpoint, and empirical convergence rates are almost exclusively measured in \textit{statistical efficiency}, i.e. n.o. iterations, as opposed to actual \textit{wall-clock time}. With new methods that potentially affect the \textit{computational efficiency}, i.e. time per iteration, such results can be delusive, with unclear usefulness in practice. Moreover, such implementations have limited capability of fine-grained exploration of aspects of synchronization mechanisms and consistency, the critical impact of which on the convergence properties has been observed analytically; 
It was shown (i)~in \cite{de2015taming} that the number of iterations until convergence increases linearly in the magnitude of the maximum staleness
and (ii)~in~\cite{alistarh2018podc} that inconsistency due to \hogwild{}-style updates further increases the same bound with a factor of $\sqrt{d}$, $d$ being the size of $\theta$. 
There is a need for further exploration of how synchronization, lock-freedom and consistency impacts the \textit{actual} wall-clock time to convergence, to facilitate work in development of standardized platforms for accelerated DL.\looseness=-1

For DL applications, convergence of sufficient quality is 
challenging to achieve, requiring exhaustive neural architecture searches and careful tuning of many \emph{hyper-parameters}. 
Unsuccessful such tuning typically results in models never converging to sufficient quality, or even executions which crash due to numerical instability in the SGD steps \cite{zhang2019empirical}.
The step size $\eta$ is among the most important hyper-parameters, while data-batch size, momentum, dropout, also play a significant role.
Tuning is vital for the convergence and end performance, and is a time-consuming process.
On one hand, parallelism in SGD is crucial for speedup, but it introduces new hyper-parameters to tune, such as number of threads, staleness bound and aspects of synchronization protocol. In addition, 
\asyncsgd{} introduces noise due to staleness, further impacting convergence and potentially causing unsuccessful executions. There is hence a need for methods enabling speedup by parallelism tolerant to existing parameters, and avoiding the overhead of tuning additional ones related to parallelism.\looseness=-1

\smallskip

\subsubsection*{Focal point and contributions} 
\begin{figure}
\centering
\includegraphics[width=0.3\textwidth]{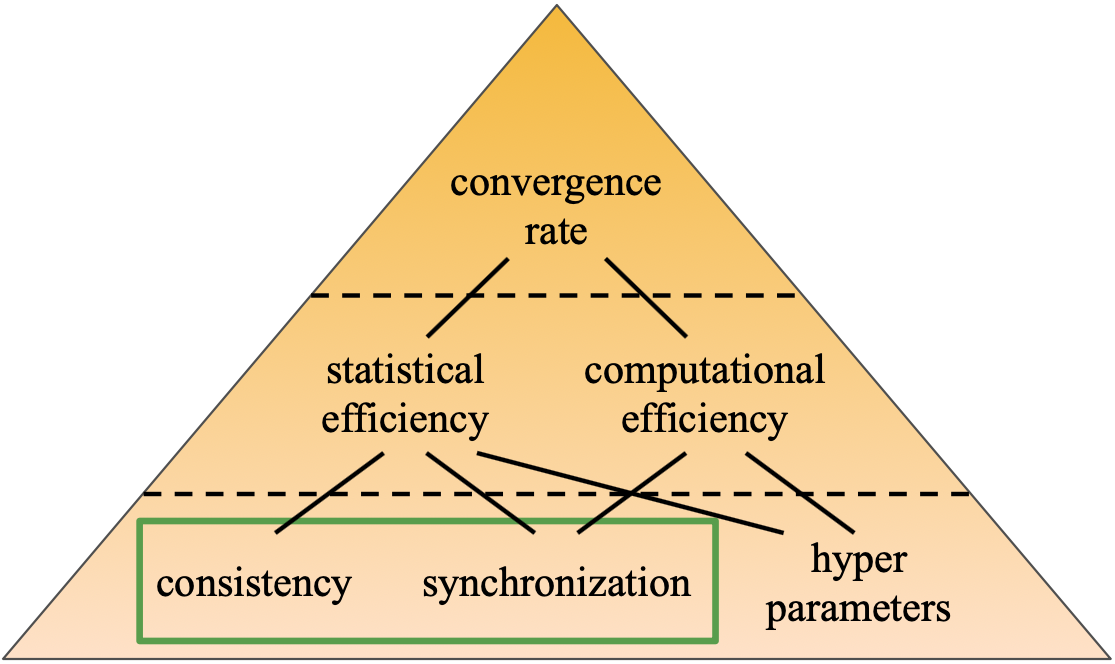}
\caption{{\small\em Convergence rate is the product of computational and statistical efficiency, sensitive to hyper-parameters tuning. We show the significant impact of
lock-free synchronization on these factors and on reducing the dependency on tuning, enabling improved convergence.\looseness=-1}}\label{fig:conv_pyramid}
\end{figure}

In summary, there are challenges in understanding the dynamics of asynchrony and consistency on the SGD convergence~\cite{wei2019automating} in practice as outlined in Fig. \ref{fig:conv_pyramid}, in particular for applications as DL.
Understanding better the tradeoff between computational and statistical efficiency is a core issue~\cite{ma2019stochastic}.
It is known that consistency helps in \asyncsgd{} \cite{alistarh2018podc}. However, whether it is worth the overhead to ensure consistency with locks or other synchronization means, to improve the overall convergence, 
is a research question attracting significant attention, as we describe here and in the related work section.

We study asynchronous SGD in a practical setting for DL. In a system-level environment, we explore aspects of synchronization, lock-freedom and consistency, and their impact on the overall convergence.
In more detail, we make the following contributions:\looseness=-1

\begin{itemize}[leftmargin=*]
    
    \item We propose \algname{}
    (\textit{\underline{l}ock-fre\underline{e}} consistent \underline{a}synchronous \underline{sh}ar\underline{ed}-memory SGD), an extensible algorithmic framework for lock-free implementations of \asyncsgd{}, allowing diverse mechanisms for consistency and for regulating contention, with efficient on-demand dynamic memory allocation and recycling.
    
    \item We analyze the proposed framework \algname{} in terms of safety, memory consumption and we introduce a model for estimating thread progression and balance in the \algname{} execution, estimating contention over time and the impact of the contention-regulation mechanism.

    \item We perform a comprehensive empirical study of the impact of synchronization, lock-freedom, and consistency on the convergence in asynchronous shared-memory parallel SGD.
    We extensively evaluate \algname{}, the standard lock-based \asyncsgd{} and its synchronization-free counterpart \hogwild{} on two DL applications, namely \emph{Multilayer Perceptrons} (MLP) and \emph{Convolutional Neural Networks} (CNN) for image classification on the image classification benchmark dataset MNIST of hand-written digits.
    We study the dynamics of contention, staleness and consistency under varying parallelism levels, confirming also the analytical observations, focusing on the \textit{wall-clock time} to convergence.
    \item We introduce a C++ framework
    supporting implementation of shared-memory parallel SGD with different mechanisms for synchronization and consistency.
    A key component is the \paramcont{} data structure, 
    providing a modularization facilitating further exploration of aspects of parallelism.
\end{itemize}

The paper is structured as follows: In section \ref{sec:prelim} we outline preliminaries and key notions for describing \algname{}, while its contention and staleness dynamics are described in sections~\ref{sec:method} and~\ref{sec:analysis}. The comprehensive empirical study is presented in \ref{sec:evaluation}, followed by further discussion of related work in section~\ref{sec:related_works}, after which we conclude in section~\ref{sec:conclusion}.\looseness=-1

\section{Preliminaries} \label{sec:prelim}

Here we give a brief background, along with a more refined description, for the questions and the metrics in focus.

\subsubsection{SGD and DL}
\textit{Artificial neural networks} (ANNs) are computational structures of simple units known as \textit{neurons}, inspired by the biological brain. Neurons are  arranged into \emph{layers}, each performing a non-linear transformation of the output from the previous layer, parameterized by a set of learnable weights. The input layer is initialized as the input to be analyzed, e.g. an image to be classified. The output layer gives the final output, e.g. the class of an image. Different types of layer arrangements give rise to a diverse class of ANN architectures, with different applications. 
Among the most prominently used are \textit{multi-layer perceptrons} (MLPs) and \textit{convolutional neural networks} (CNNs) \cite{baldominos2019survey}, where MLPs consist of layers \textit{densely} connected through a weight matrix, and CNNs of sparsely connected performing filter convolutions, used in conjunction with \textit{MaxPool} downsampling layers.
Some more information on MPLs and CNNs appears in the \suppl{}.\looseness=-1

The aforementioned weights and filters consist of parameters, learned through the training process. 
We refer to the collection of all such parameters belonging to an ANN, flattened into a 1D array, as the \textit{parameter vector}, denoted as $\theta_t$, at iteration $t$ of SGD. 
This abstraction is used in subsequent sections when arguing regarding consistency and progress. Non-linear activation functions are applied after each layer, where common choices are the \textit{ReLU} function $\sigma(x) = max(0,x)$ for all layers except the last, where instead the \textit{softmax} activation function $\sigma_i(x) = e^{x_i}/\sum^{|x|}_{j=1} e^{x_j}$, for each output neuron $i$, is used in order to acquire a predicted probability distribution. With this, an error measure $f(\theta)$ can be defined, the minimization of which constitutes the training process.\looseness=-1

The \emph{metrics} of interest 
are (i) \textit{statistical efficiency}, i.e. the number of SGD iterations required until reaching an error threshold $f(\theta^*)<\epsilon$, i.e. $\epsilon$-convergence (ii) \textit{computational efficiency} measuring the wall-clock time per iteration and, most importantly (iii) the \textit{overall convergence rate}, i.e. the wall-clock time until $\epsilon$-convergence, of most relevance in practice.\looseness=-1

\subsubsection{System Model} 
We consider a system with $m$ concurrent asynchronous threads, 
with access to shared memory through atomic operations to read, write and read-modify-write, e.g. CompareAndSwap (CAS), FetchAndAdd (FAA) \cite{herlihy2011art} on single-word locations. 
Each thread $A$ computes SGD updates (\ref{eq:sgd_step}) according to a pre-defined algorithm, in the context outlined in the previous paragraphs. 
Since $A$ must read the current state $\theta_t$ prior to computing the corresponding stochastic gradient $\nabla f(\theta_t)$, before $A$'s updates take place, there can be intermediate, referred to as \textit{concurrent updates}, from other threads, 
The number of such updates, between $A$'s read of the $\theta_t$ vector and $A$'s update to apply its calculated gradient $\nabla f(\theta_t)$, defines the \emph{staleness} $\tau$ of the latter update. When there is lack of synchronization, as in \hogwild{}, a total order of the updates is not imposed, and 
the definition of the staleness of an update is not straightforward; we adopt a definition similar to  \cite{alistarh2018podc}.
We refer to Section (\ref{sec:method}) for details on how the staleness is calculated for the different algorithms, and thereby the total order of the updates.
Under the system model above, we have that the asynchronous SGD updates according to (\ref{eq:sgd_step}) instead will follow\looseness=-1
\begin{align}
    \theta_{t+1} \leftarrow \theta_t - \eta \nabla f(v_t) \label{eq:asynch-sgd}
\end{align}
where $v_t = \theta_{t-\tau_t}$ is the thread's \textit{view} of $\theta$.

\subsubsection{Synchronization methods and consistency}
For consistency on concurrently accessed data, different methods for thread synchronization exist, the most traditional one being \emph{locks} for mutually exclusive access.
\emph{Non-blocking synchronization} avoids the use of locks. \cite{herlihy2011art}. 
A common choice is  \emph{lock-free synchronization}, ensuring that in the presence of concurrent object accesses, some are able to complete in a bounded number of steps, thus guaranteeing system progress.
Such synchronization mechanisms usually implement a \emph{retry loop} involving CAS or equivalent, in which a thread might need to repeat, in  case  another thread has succeeded.

Besides \emph{progress} guarantees, to argue about concurrent data  accesses, we consider \emph{data consistency}. The most common is \emph{atomicity} (aka \emph{linearizability}, with non-blocking synchronization), and it implies that concurrent object operations  act as if they are executed in sequence, affecting state and returning values according to the object's sequential specification~\cite{herlihy2011art}.
\looseness=-1

\subsubsection{Problem overview}
In the following, we focus on exploring the effectiveness of asynchronous parallel algorithms for SGD, for training \textit{deep neural networks} (DNNs). We study the computational and statistical efficiency for different applications, and the overall time to $\epsilon$-convergence. We explore in particular the effect of different synchronization mechanisms on consistency, contention and staleness, and the resulting impact on the convergence and memory consumption. \looseness=-1

\section{The \algname{} framework} \label{sec:method}

In the following we define \algname{} 
along with the proposed \paramcont{}  data structure's common interface, containing the values of the parameter vector, as well as metadata used for memory recycling.
We also express \asyncsgd{} and \hogwild{} using this interface; both are well established versions of parallel SGD implementations~\cite{recht2011hogwild, agarwal2011distributed}. Modified versions, optimized for specific applications, have been proposed, e.g. in \cite{zhang2016hogwildpp}, however not in the context of DL.
In the following, we use 
them as general baselines, representative of the classes of \textit{consistent} asynchronous SGD algorithms and the \textit{synchronization-free, inconsistent} \hogwild{}-style ones.

\subsubsection{Introducing \paramcont}
Considering (\ref{eq:sgd_step}), each worker in parallel SGD reads the shared data object $\theta$, computes a gradient and updates the former. 
We propose a set of core components for this type of data structure, \paramcont{}, providing possibilities to get parameter values and submit updates.
An instantiation of \paramcont{} can be local or shared among threads.
For concurrent accesses to it, its implementation can provide certain consistency and progress guarantees (cf. section~\ref{sec:prelim}).
Hence studying shared memory data-parallel SGD implementations with synchronization in focus, is to study implications of the properties of the algorithmic implementations of the parameter vector seen as shared object, connecting to and extending work in the literature on bulk operations on container data structures~\cite{nikolakopoulos2015consistency}.\looseness=-1

\SetKwProg{Fn}{Function}{:}{}

\begin{algorithm}

    \scriptsize

    Float[d] $theta$  // vector of dimension $d$ \\
    Int $t \leftarrow 0$ \hfill // sequence number of the most recent update of theta \\
    Int $n\_rdrs \leftarrow 0$ \\
    Bool $stale\_flag \leftarrow false$, $deleted \leftarrow false$  \\
    
    \vspace{1pt}
    
    \SetKwFunction{randinit}{rand\_init}
    \Fn{\randinit{}}{
        $theta \leftarrow \mathcal{N}(0,0.01)$
    }
    
    \SetKwFunction{safedelete}{safe\_delete}
    \Fn{\safedelete{}}{
        \uIf{$stale\_flag \land n\_rdrs = 0 \land CAS(deleted, false, true)$}{
            \textbf{delete} $theta$ \label{line:safe_delete}
        }
    }
    
    \SetKwFunction{startreading}{start\_reading}
    \Fn{\startreading{}}{
        $param.n\_rdrs.fetch\_add(1)$
    }
    
    \SetKwFunction{stopreading}{stop\_reading}
    \Fn{\stopreading{}}{
        $n\_rdrs.fetch\_add(-1)$ \\
        $self.\safedelete{}$
    }
    
    \SetKwFunction{PCupdate}{update}
    \Fn{\PCupdate{$\delta$, $\eta$}}{
        $t.fetch\_add(1)$ \\
        \For{$i=0,\dots,d-1$}{
            $theta[i] \leftarrow theta[i] - \eta \cdot \delta[i]$
        }
    }
    
    \caption{\paramcont{} core components}
    \label{algorithm:PC_implementation}

\end{algorithm}

Algorithm~\ref{algorithm:PC_implementation} describes the core components for the algorithmic implementation of \paramcont{}. A main one is the array $theta$ of dimension $d$ (typically a very large number in DL applications, e.g. in the well-known AlexNet\cite{krizhevsky20122012} CNN architecture there are 62,378,344 parameters).
A read of the parameters can be accomplished by getting a pointer to $theta$, while function \PCupdate{} performs the addition (\ref{eq:asynch-sgd}) on $theta$.
Notice that algorithm~\ref{algorithm:PC_implementation} does not provide specific synchronization for protecting reads of updates, which is instead left to the algorithmic implementation's \emph{``front-end"} to specify, depending on the demands of consistency. It provides however additional methods and metadata  for keeping track of accesses and for recycling memory, as explained further in this section.
While there is some resemblance with a multi-word register~\cite{larsson2004multi,ianni2018anonymous}, two significant issues here are (i)~the nature of the update, which is a bulk  Read-Modify-Write operation and (ii)~the very large value of $d$, posing challenges both from the memory and from the timing (retry loop size) perspectives.\looseness=-1

\begin{algorithm}
    \scriptsize

    GLOBAL ParamVector $PARAM$ \hfill \\
    GLOBAL Float $\eta$ \hfill // step size \\
    GLOBAL Lock $mtx$ \hfill // for accessing shared parameters
    
    \underline{Initialization}\;
    
    $PARAM \leftarrow $ new $ParamVector()$ \\
    $PARAM.\randinit{}$ \hfill // randomly initialize parameters \\
    
    \underline{Each thread}\;
    
    $local\_grad \leftarrow $ new $ParamVector()$ \hfill // local gradient memory \\
    $local\_param \leftarrow $ new $ParamVector()$
    
    \Repeat {convergence}{
        $mtx$.lock() \\
        $local\_param.theta = $ copy$(PARAM.theta)$ \\
        $mtx$.unlock() \\
        $local\_grad.theta \leftarrow comp\_grad(local\_param.theta)$ \\
        $mtx$.lock() \\
        $PARAM$.\PCupdate{$local\_grad.theta, \eta$} \\
        $mtx$.unlock() \\
    }
    
    \caption{\asyncsgd{}}
    \label{algorithm:async_implementation}
\end{algorithm}

\subsubsection{Baselines outline}
Algorithm~\ref{algorithm:async_implementation} shows the lock-based \asyncsgd{}, one of the baselines, achieving consistency in the reads and the updates of the parameters through locking. This introduces an overhead, influencing the thread interleaving, with unclear implications on staleness and statistical efficiency. This is further explored in Section \ref{sec:evaluation}. There is one shared variable of type \paramcont{}, $PARAM$, and  two local ones to each thread, one with a copy of the latest state of the shared parameter vector ($local\_param$) and one for storing the gradient ($local\_grad$).
\hogwild{}'s algorithmic implementation is similar to Algorithm~\ref{algorithm:async_implementation}, except that the locks are removed, since no synchronization happens among the threads accessing the parameter vector. Certain overhead is thus eliminated, however at the cost of inconsistency in the parameter updates. 
The algorithm outline is available in the \suppl{}.
For problems with sparse gradients the lack of synchronization will not significantly impact the convergence, since the \PCupdate{} operation will only influence a few of the $d$ components in $theta$.
For DL applications though, its influence is not well understood.
\looseness=-1

\subsubsection{\algname{}: Lock-free consistent \asyncsgd{}}

\begin{algorithm}

    \scriptsize

    GLOBAL ParamVector ** $P$ \hfill // address to latest pointer (cf. Fig.~\ref{fig:algos})\\
    GLOBAL Float $\eta$ \hfill // step size \\
    GLOBAL Int $T_p$ \hfill // persistence threshold \\
    
    \vspace{1pt}
    
    \SetKwFunction{latestpointer}{latest\_pointer}
    \Fn{\latestpointer{}}{
        \Repeat{break}{  
            $latest\_param \leftarrow *P$ \hfill // fetch latest pointer \\
        
            $latest\_param.\startreading{}$ \hfill // prevent it from  recycling   \\
            
            \uIf{$\neg latest\_param.stale\_flag$}{ \label{line:staleness_check}
                \textbf{return} $latest\_param$
            }
            \Else{
                $latest\_param.\stopreading{}$ \hfill // avoid returning stale vector, let it be recycled and repeat to get a fresher one 
            }
        }
    }
    
    \underline{Initialization}\; 
    
    $init\_pv \leftarrow $ new $ParamVector()$ \hfill // pointer to initial parameters \\
    $init\_pv.\randinit{}$ \hfill // randomly initialize parameters \\
    $P \leftarrow \& init\_pc$ \hfill // address of initial pointer \\
    
    \vspace{1pt}
    
    \underline{Thread $i$}\;
    
    $local\_grad \leftarrow $ new $ParamVector()$ \hfill // local gradient memory \\
    
    \Repeat{convergence}{
    
        $latest\_param \leftarrow \latestpointer{}$ \\
        
        $local\_grad.theta \leftarrow comp\_grad(latest\_param.theta)$ \\
    
        $latest\_param.\stopreading{}$ \\
        
        $new\_param \leftarrow $ new $ParamVector()$ \hfill // new parameters
        
        Int $num\_tries \leftarrow 0$ \hfill // prepare for the \casretryloop{} loop \\
    
        \Repeat{$succ$}{  
    
            $latest\_param \leftarrow \latestpointer{}$ \\
            
            $new\_param.theta \leftarrow $ copy$(latest\_param.theta)$ \\
            
            $new\_param.t \leftarrow latest\_param.t$
            
            $latest\_param.\stopreading{}$ \\
    
            $new\_param$.\PCupdate{$local\_grad.theta$, $\eta$}  \\
            
            $succ \leftarrow CAS\big(P, latest\_param, new\_param\big)$ \label{line:CAS} \\
            
            \uIf{succ}{
                $latest\_param.stale\_flag \leftarrow true$ \\
                $latest\_param$.\safedelete{}
            }
            \Else{
                $num\_tries \leftarrow num\_tries + 1$ \\
                \uIf{$num\_tries > T_p$}{
                    \textbf{delete} $new\_param$ \\
                    \textbf{break}
                }
            } 
        } % // LAU-SPC (LoadAndUpdate-StorePersistenceConditional) loop
    }
    
    \caption{\algname{}}
    \label{algorithm:leashed_implementation_successor}
\end{algorithm}

\begin{figure}[t]
    \centering
    \includegraphics[width=0.3\linewidth]{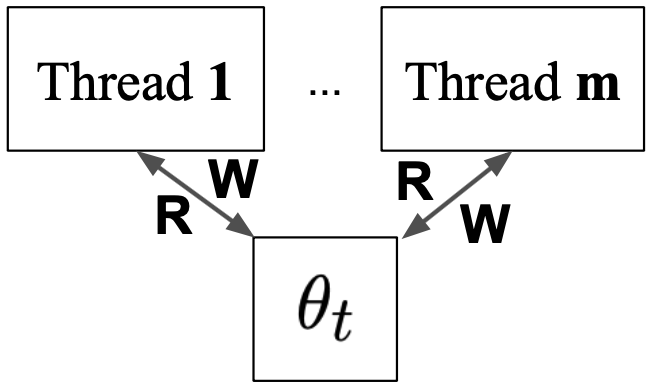}
    \quad
    \includegraphics[width=0.4\linewidth]{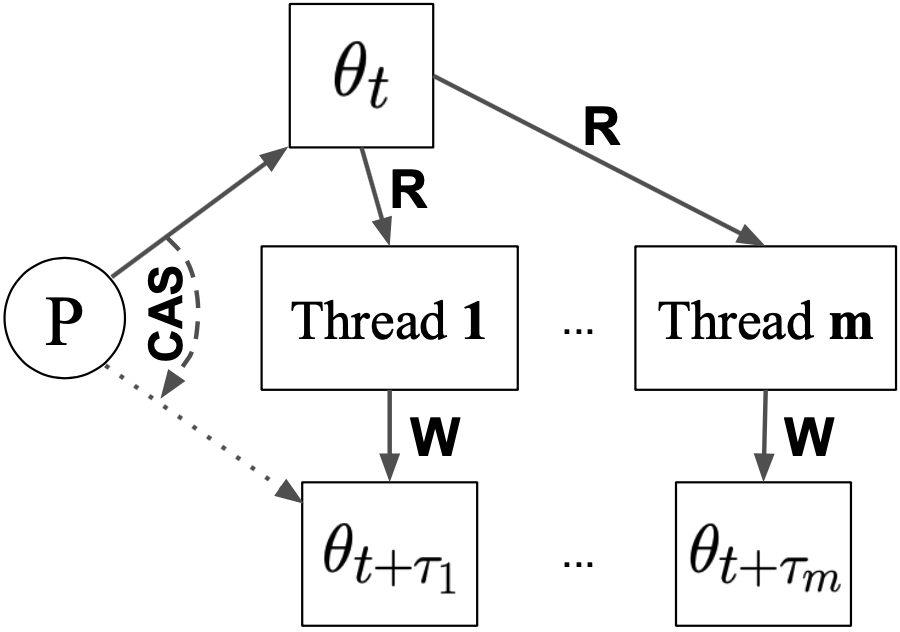}
    \caption{{\small\em Illustration of data access in \asyncsgd{} and \hogwild{} (\textit{left}) and \algname{} (\textit{right}). For \asyncsgd{} the read and write operations are protected through mutual exclusion. For \algname{}, each thread accesses $\theta_t$ only through a read operation, then computes the update at a new memory location, becoming a candidate for~$\theta_{t+\tau}$}.\looseness=-1}
    \label{fig:algos}
\end{figure}

The key points and arguments supporting \algname{}, which is shown in pseudocode in Algorithm~\ref{algorithm:leashed_implementation_successor}, using \paramcont{} core components from Algorithm~\ref{algorithm:PC_implementation}, are as follows:\\   
    \emph{{\bf P1.} Local calculation and sharing of new parameter values:} Each thread manages its update locally $new\_param$, and attempts to publish the result in a single atomic CAS operation (line \ref{line:CAS}), switching a global pointer $P$ to point to its new instance (Fig.~\ref{fig:algos}). 
    As a successful CAS replaces the previous ``global" vector, copies of parameter vectors that become global are \emph{totally ordered} on their sequence number, $t$. 
    A vector that has been replaced using the aforementioned CAS, is labeled as \emph{stale} through a boolean flag ($stale\_flag$ in \paramcont{}) that is one of the data structure's fields.\\
    \emph{{\bf P2.} Memory recycling:} Since a new \paramcont{} is needed for each such update, a simple yet efficient \emph{recycling} mechanism of \emph{stale and unusable} ones ensures that the memory used is bounded. Besides the label for marking a \paramcont{} instance as stale (ensuring no new readers, making it a candidate for recycling), the field $n\_rdrs$, indicates whether the \paramcont{} should persist due to active readers.\looseness=-1\\
    \emph{{\bf P3.} 
    Lock-free atomic reads of the shared vector:} To access the global \paramcont{} threads acquire a \textit{pointer} to the most recent by accessing $P$. Through that pointer, the thread can access and use the $theta$ and metadata of that \paramcont, in particular for calculating the gradient without copying. While a \paramcont{} $V$ is in use, 
    $V.n\_rdrs$ is non-zero (it is atomically increment-able and decrement-able in the \startreading{} and \stopreading{} functions). 
    Note that the update of the global pointer $P$, and the marking of the previous global vector as stale, are two operations. Hence, for a thread to acquire the latest \paramcont{} in a concurrency-safe manner, this must be done in a retry loop, in \latestpointer{}. Due to this fact and how the global pointers are updated, a read preceded by another read will not return parameter values older than its preceding read returned.\\
    \emph{{\bf P4.} Conditions for safe recycling:} For reclaiming the memory of a \paramcont{} $V$, the $V.stale\_flag$ must be $true$ and $V.n\_rdrs$ must be zero. The first condition ensures that the \paramcont{} instance is not the most recently published, and its address is no longer available to any thread (Algorithm \ref{algorithm:leashed_implementation_successor}, line \ref{line:CAS}), ensuring no additional future accesses. The second condition ensures that no thread is currently accessing $V$, with the exception when a thread just acquired a pointer that just became stale, which subsequently will repeat after the staleness check that follows in line~\ref{line:staleness_check}. Note that stale instances of \paramcont{} will be reclaimed by the last thread to access it, when calling \stopreading{}.\looseness=-1 \\
    \emph{{\bf P5.} Lock-free atomic updates of the shared vector:} The publish is attempted through a CAS invoked in a retry loop, and if it fails, another thread must have succeeded. 
    Update attempts are repeated until CAS succeeds, or until a persistence bound $T_p$ decided by the user has been exceeded.
    The loop thus implies lock-free progress guarantees. 
    For $T_p=0$ it implies similar semantics as the LoadLinked/StoreConditional primitive, hence its name  \emph{LoadAndUpdate}-\emph{StorePersistenceConditional} (\casretryloop{}).
    Note that bounded $T_p$ essentially implies bounded retries.
    As formulated in (\ref{eq:asynch-sgd}), due to asynchrony, the gradients can be applied on a different \paramcont{} instance than the one that was used to compute the gradient. Hence, after finishing the gradient computation, threads acquire the pointer to the most recent published \paramcont{} instance a second time (Figure~\ref{fig:algos}), on which the update will be applied. The result is then a candidate for publishing, the success of which is decided as described above, implying update atomicity.

Based on the previous paragraphs, (in particular on points P1, P3 and P5, respectively points P2 and P4) we have:

\begin{lemma}  
Reads and updates of the $\theta$ vector by \algname{}, \latestpointer{} function and \casretryloop{} loop, satisfy lock-freedom and atomicity.
\end{lemma}

\begin{lemma} \label{lemma:safe_mem_rec}
The memory recycling in \algname{}
(i) is safe, i.e. will not reclaim memory which can be used by any thread for reading or updating and (ii) bounds the memory to max $3m$ \paramcont{} instances simultaneously.
\end{lemma}

\subsubsection*{A note on memory consumption}
Note that \asyncsgd{} and \hogwild{} need $2m +1$ instances of \paramcont{} constantly. In \algname{} threads compute gradients based on a published \paramcont{} instance, which will never be altered by any thread. After the gradient computation is finished, additional memory is allocated for new parameters. This mechanism enables an overall reduced memory footprint, in particular when gradient computation is time consuming.
This is confirmed empirically in section~\ref{sec:evaluation}.

\section{Contention and staleness } \label{sec:analysis}

In the following we analyze the dynamics and balance of the proposed \algname{}, the effect of the \textit{persistence bound}, and its impact on the contention and staleness.

\subsubsection{Dynamics of \algname{}}
We analyze the dynamics of the threads, their progression under concurrent execution of \algname{}. The model is similar to a G/G/1 queue, but with arrival and departure rates $\lambda_t, \mu_t$ varying over time, depending on the current state of the system.

    \begin{table*}[!b]
    \caption{Summary of experiments}
    \begin{center}
    \begin{tabular}{|c|c|c|c|c|c|c|}
    \hline
    &\multicolumn{6}{|c|}{\textbf{Experiment overview}} \\
    \cline{2-7}
    \textbf{Step} & \textbf{\textit{Architecture}}& \textbf{\textit{Description}}& \textbf{\textit{N.o. threads $m$}} & \textbf{\textit{Precision $\epsilon$}} & \textbf{\textit{Step size $\eta$}} & \textbf{\textit{Outcome}} \\
    \hline
    \textbf{S1} & MLP$^*$& Hyper-parameter selection & 1-68 & $50\%$ & $0.01-0.09$ & Fig. \ref{fig:convrate_compeff} \\
    \textbf{S2} & MLP$^*$& High-precision convergence & 16 & $50\%, 10\%, 5\%, 2.5\%$ & $0.05$ & Fig. \ref{fig:precision_MLP}-\ref{fig:taudist_MLP} \\
    \textbf{S3} & CNN$^*$ & Convergence rate & 16 & $75\%, 50\%, 25\%, 10\%$ & $0.05$ & Fig. \ref{fig:cnn_tests} \\
    \textbf{S4} & MLP$^*$ & High parallelism & 24, 34, 68 & $75\%, 50\%, 25\%, 10\%$ & $0.05$ & Fig. \ref{fig:precision_MLP}-\ref{fig:taudist_MLP} \\
    \textbf{S5} & MLP$^*$, CNN$^*$ & Memory consumption & 16, 24, 34 & any & $0.05$ & 
    %Fig. \ref{fig:mem_con}
    presented in text$^*$
    \\
    \hline
    \multicolumn{4}{l}{$^*$Details appear in the \suppl{}}
    \end{tabular}
    \label{tab:summary_exp}
    \end{center}
    \end{table*}
    
For a single thread executing the gradient computation, the rate of arrival to the \casretryloop{} (retry) loop is $\lambda^{(1)} = 1 / T_c$, where $T_c$ is the gradient computation time. For an $m$-thread fully concurrent execution, the arrival rate scales proportionally to the number of threads currently outside the \casretryloop{} loop, hence $\lambda^{(m)} = (m-n) \lambda^{(1)}$ where $n$ denotes the number of threads in the retry loop. Similarly, for the departure rate from the \casretryloop{} loop we have $\mu^{(1)} = 1/T_u$ where $T_u$ is the execution time of the \paramcont{} \PCupdate{}. In summary: \looseness=-1
\begin{align}
    \lambda_t^{(m)} = \frac{m-n_t}{T_c}, \ \  %\label{eq:arrival_rate} 
    \mu_t = \frac{n_t}{T_u} \label{eq:arrival_departure_rate}
\end{align}
We then describe the dynamics of how threads enter and leave the \casretryloop{} retry loop of \algname{} as follows:
\begin{align}
    n_{t+1} = n_t + \frac{m-n_t}{T_c} - \frac{n_t}{T_u} \label{eq:leashed_dynamics}
\end{align}
where $n_t$ is the number of threads executing the retry loop at time $t$. Note that the system (\ref{eq:leashed_dynamics}) has a fixed point $n^* = (T_c/T_u + 1)^{-1} m$ at which the number of threads in the retry loop will stay constant. Note that $n^*$ rewrites to $n^*/m = T_u / (T_u+T_c)$, i.e. that thread balance at the fixed point depends solely on the relative size of the update time $T_u$, highlighting the importance of the ratio $T_u/T_c$. In section \ref{sec:evaluation} we show closer measurements of $T_c, T_u$ for different applications.

In the following, we study how $n_t$ progresses for \algname{}, stability and convergence about the fixed point.
\begin{theorem} \label{theorem:n_t}
    Assume we have an $m$-thread system where threads arrive to and depart from the \algname{} \casretryloop{} loop with the rates in (\ref{eq:arrival_departure_rate}). Then, we have that the number $n_t$ of threads in the retry-loop at time $t$ is given by
    \begin{align}
        n_t = \frac{1 - (1-T_c^{-1}-T_u^{-1})^t}{1+T_c/T_u} m + (1-T_c^{-1}-T_u^{-1})^t n_o \label{eq:n_t}
    \end{align}
    where $T_c, T_u$ denotes the time for gradient computation and update, and $n_0$ is the initial number of threads in \casretryloop{}.
\end{theorem}
Due to space constraints, the proof appears in the \suppl{}.
\begin{corollary}
    The fixed point $n^*$ is stable, and the system will converge towards $\lim_{t\to\infty} n_t = n^*$
    for any initial $n_0$.
\end{corollary}
The result is confirmed by taking $t \rightarrow \infty$ in (\ref{eq:n_t}).

The above results enable understanding of the dynamics of how threads progress throughout the execution, in particular that they converge to a balance between gradient computation and the \casretryloop{}, which will be used in the following.

\subsubsection{Persistence analysis} 
The persistence bound implies a threshold on the maximum number of failed CAS attempts in \algname{}, before threads compute a new gradient. This implies an increase, denoted by $\gamma>0$, in departure rate from the \casretryloop{} retry loop, proportional to the number of threads currently in the retry loop as follows:
\begin{align}
    \mu_t &= \frac{n_t}{T_u} ( 1 + \gamma ) \label{eq:deprate_pers}
\end{align}

\begin{corollary}
    Under the same conditions as in Theorem \ref{theorem:n_t}, but using the departure rate (\ref{eq:deprate_pers}), the fixed point moves to
    \begin{align}
        n^*_\gamma = \big( \frac{T_c}{T_u}(1+\gamma) + 1 \big)^{-1} m
    \end{align}
\end{corollary}
Note that (i) $n^*_\gamma < n^*$ and (ii) $n^*_\gamma$ vanishes as $\gamma$ grows, showing the contention-regulating capability through a persistence bound, i.e. an increased $\gamma$.

As pointed out in \cite{backstrom2019mindthestep}, the complete staleness $\tau_t$ of an update $\nabla f(v_t)$ according to (\ref{eq:asynch-sgd}) is comprised of two parts: $\tau_t = \tau_t^c + \tau_t^s$
where $\tau_t^c$ counts the number of published updates concurrent to the computation of $\nabla f(v_t)$, and $\tau_t^s$ counts the ones that compete with the update in focus and are scheduled before it; in particular here, the latter counts the competing updates in 
 the \casretryloop{} loop that succeed before that update. 
 %Assuming now the reasonable estimation 
 Considering now the estimation $\textbf{E}[\tau_t^s] \approx n^*_\gamma$, it follows that the persistence mechanism described above for reducing contention effectively regulates the additional staleness component due to scheduling of ready gradients.

E.g., consider $T_p=0$: for each published update there was no failed CAS, hence no other update was published after the corresponding gradient was used. Then $\tau_t^s = 0$, which is the maximum staleness reduction possible here.
%with this regulation method. 
In section~$\ref{sec:evaluation}$ we study this empirically, showing it holds in practice and is effective for regulating contention and tune the staleness. \looseness=-1

\section{Evaluation} \label{sec:evaluation}

We present the results from our extended empirical study, benchmarking the methods in Section \ref{sec:method}, studying influence of consistency and associated synchronization, on the metrics described in Section~\ref{sec:prelim}: convergence rate, statistical and computational efficiency, and memory consumption.The algorithms included are sequential SGD (SEQ), Lock-based \asyncsgd{} (ASYNC), \hogwild{} (HOG), and \algname{} with persistence $\infty, 1, 0$ (LSH\_ps$\infty$, LSH\_ps1, LSH\_ps0). \looseness=-1

\begin{figure*}
\centering
\begin{minipage}[b]{.62\textwidth}
\includegraphics[width=\textwidth]{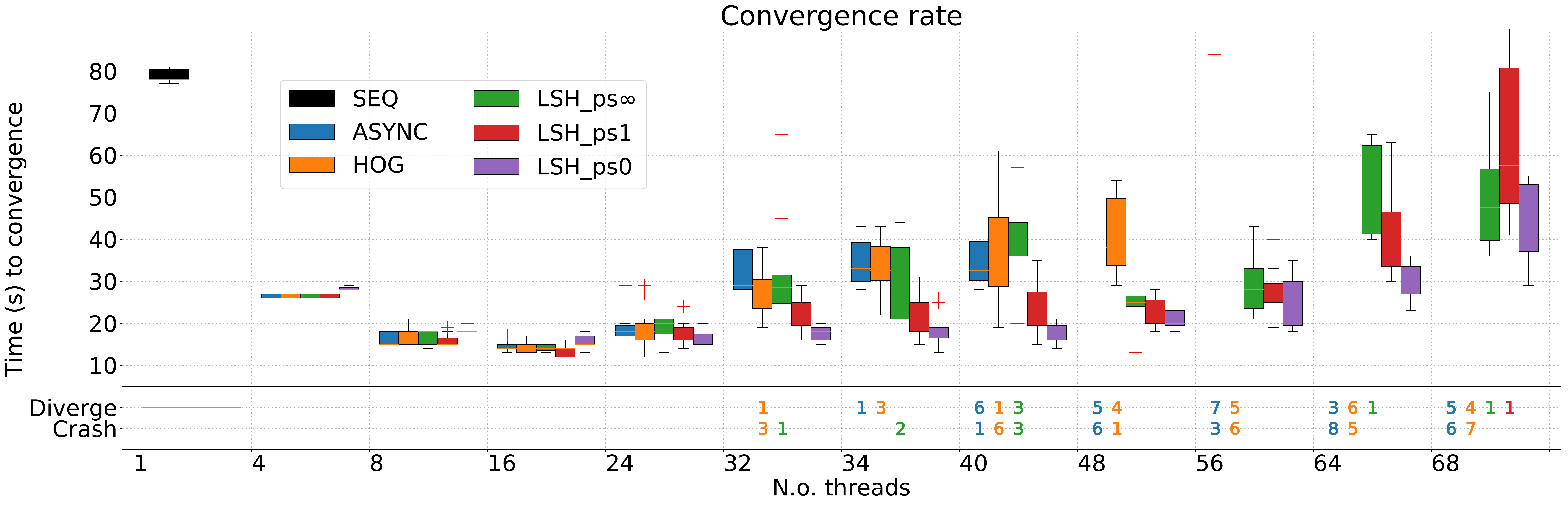}
\end{minipage}
\begin{minipage}[b]{.37\textwidth}
\includegraphics[width=\textwidth]{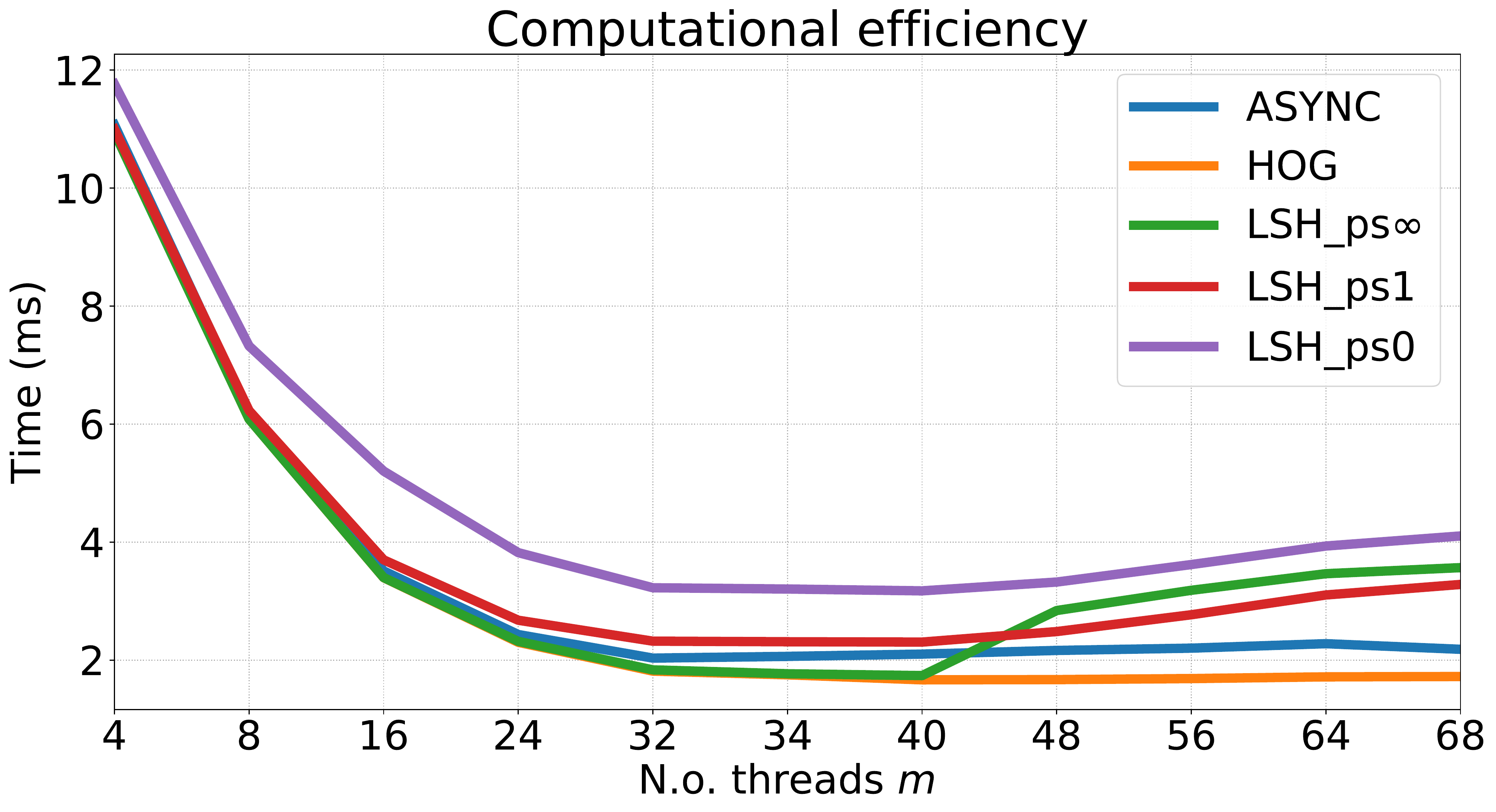}
\end{minipage}
\caption{
\textit{Left:} $\epsilon$-convergence rate for MLP training under varying parallelism, measuring wall-clock time until reaching an error threshold ($\epsilon=50\%$ of initial error, providing a comparison of the general scalability). The optimum for the baselines ($m=16$) is used in subsequent tests for a fair comparison. Under increased parallelism ($m>16$), the convergence rate for the baselines (ASYNC, HOG) deteriorates, with many unstable executions, never achieving $\epsilon$-convergence due to the instability from increased staleness. The proposed framework ($LSH\_psX$, persistence bound $X$) on the other hand provides stable and fast convergence for up to 56 threads, with minimal penalty from staleness. \textit{Right:} Computational efficiency, i.e. wall-clock computation time per SGD iteration. Computation time remains constant for the baselines under higher parallelism, although the many executions fail completely to converge, and would be wasted time in practice. The self-regulative property of \algname{} on the other hand increases the computation time moderately under high parallelism, balancing latency and throughput under contention, and can hence achieve stable convergence in far more instances.}\label{fig:convrate_compeff}
\end{figure*}

\begin{figure*}
\centering
\begin{minipage}[b]{.32\textwidth}
\includegraphics[width=\textwidth]{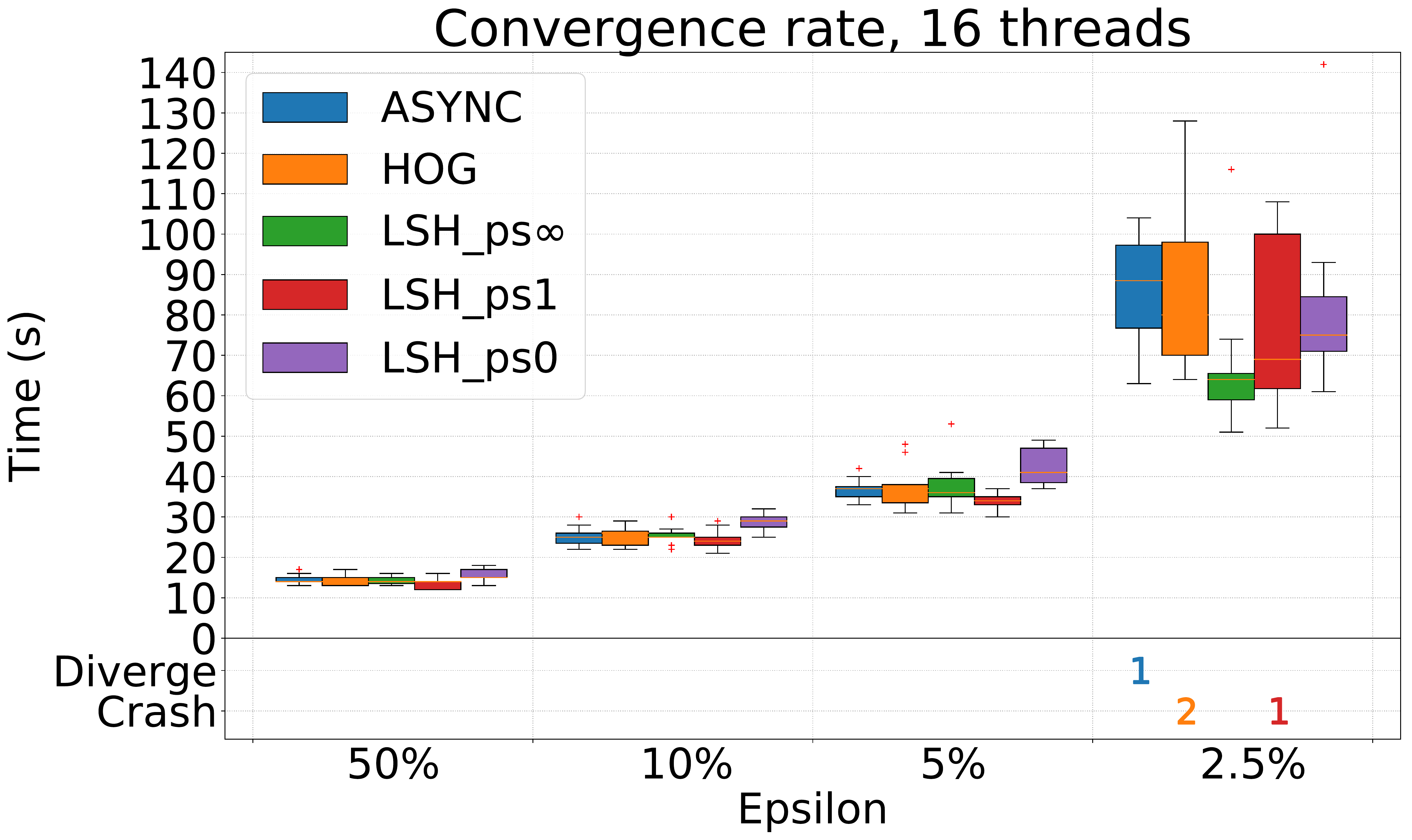}
\end{minipage}
\begin{minipage}[b]{.32\textwidth}
\includegraphics[width=\textwidth]{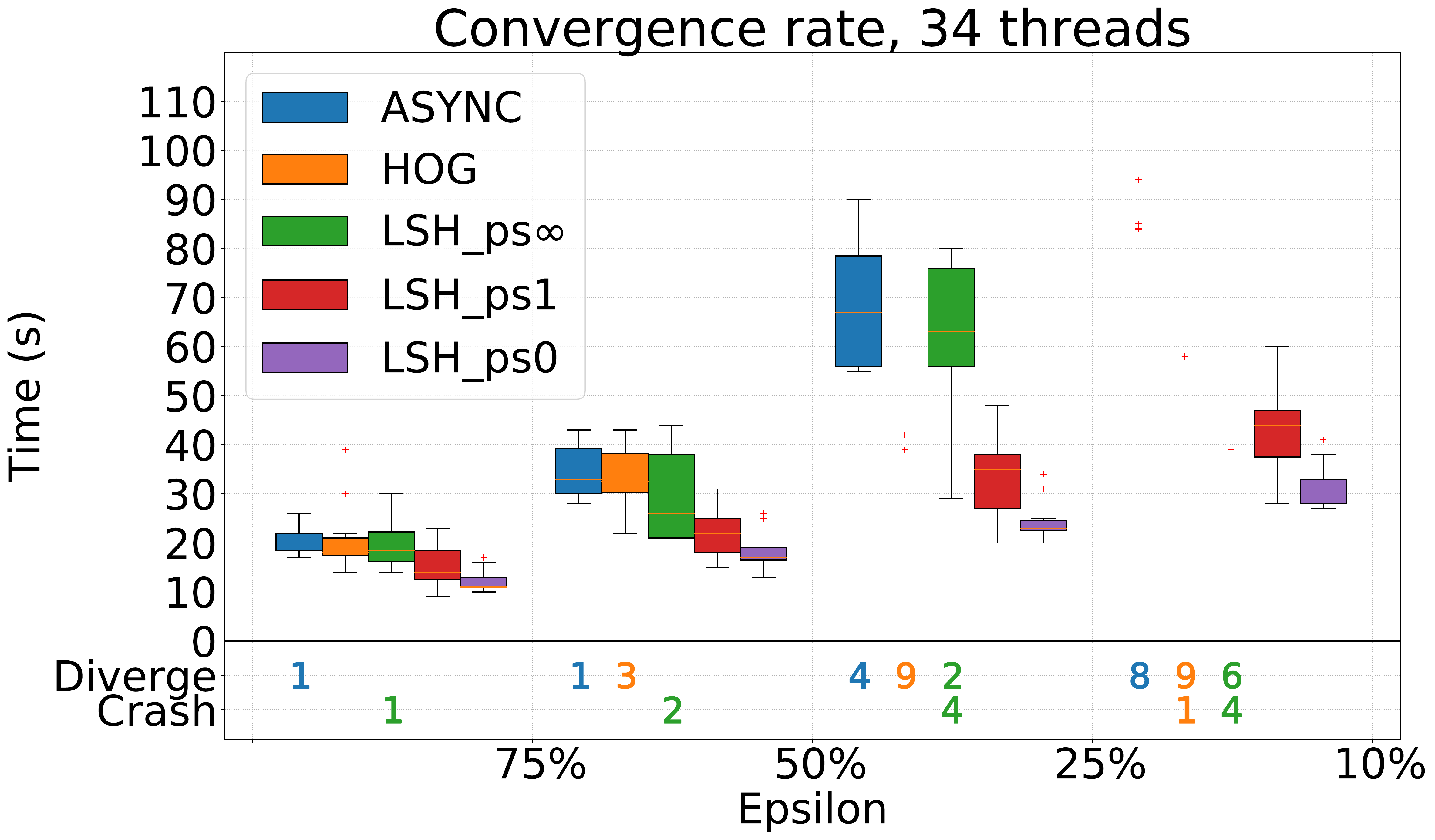}
\end{minipage}
\begin{minipage}[b]{.32\textwidth}
\includegraphics[width=\textwidth]{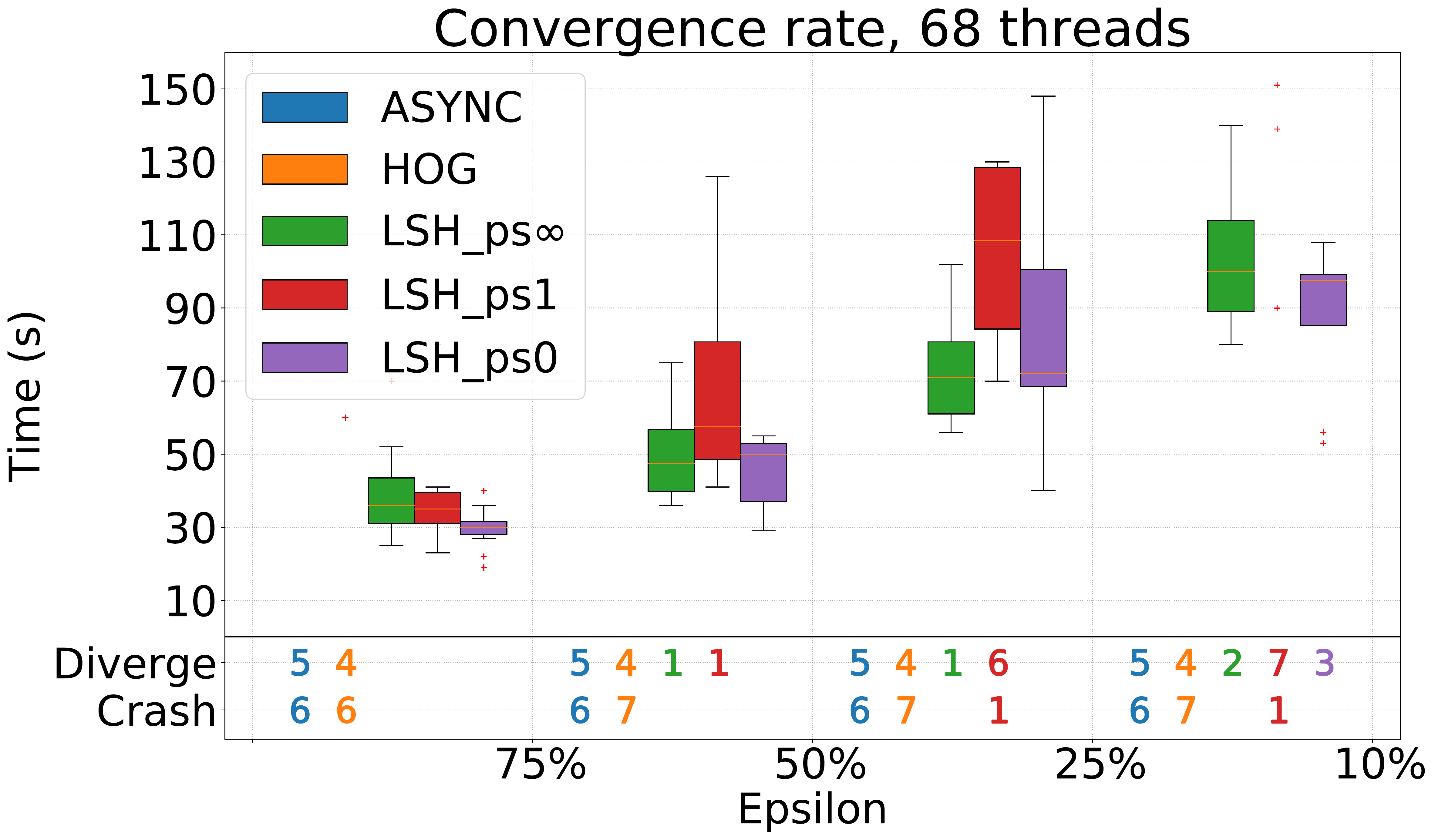}
\end{minipage}
\caption{$\epsilon$-convergence rate for MLP with $m=16$ threads to high precision (\textit{left}), $m=34$ threads (\textit{middle}) and maximum parallelism $m=68$ threads (\textit{right}). The baselines (ASYNC, HOG) show an overall slower convergence and higher number of executions that 
%crash or diverge 
fail before reaching the high precision (e.g. $\epsilon=10\%$), especially under maximum parallelism $m=68$, where no baseline execution managed to reach $\epsilon=50\%$ of the error at initialization.}\label{fig:precision_MLP}
\end{figure*}

\begin{figure*}
\centering
\begin{minipage}[b]{.31\textwidth}
\includegraphics[width=\textwidth]{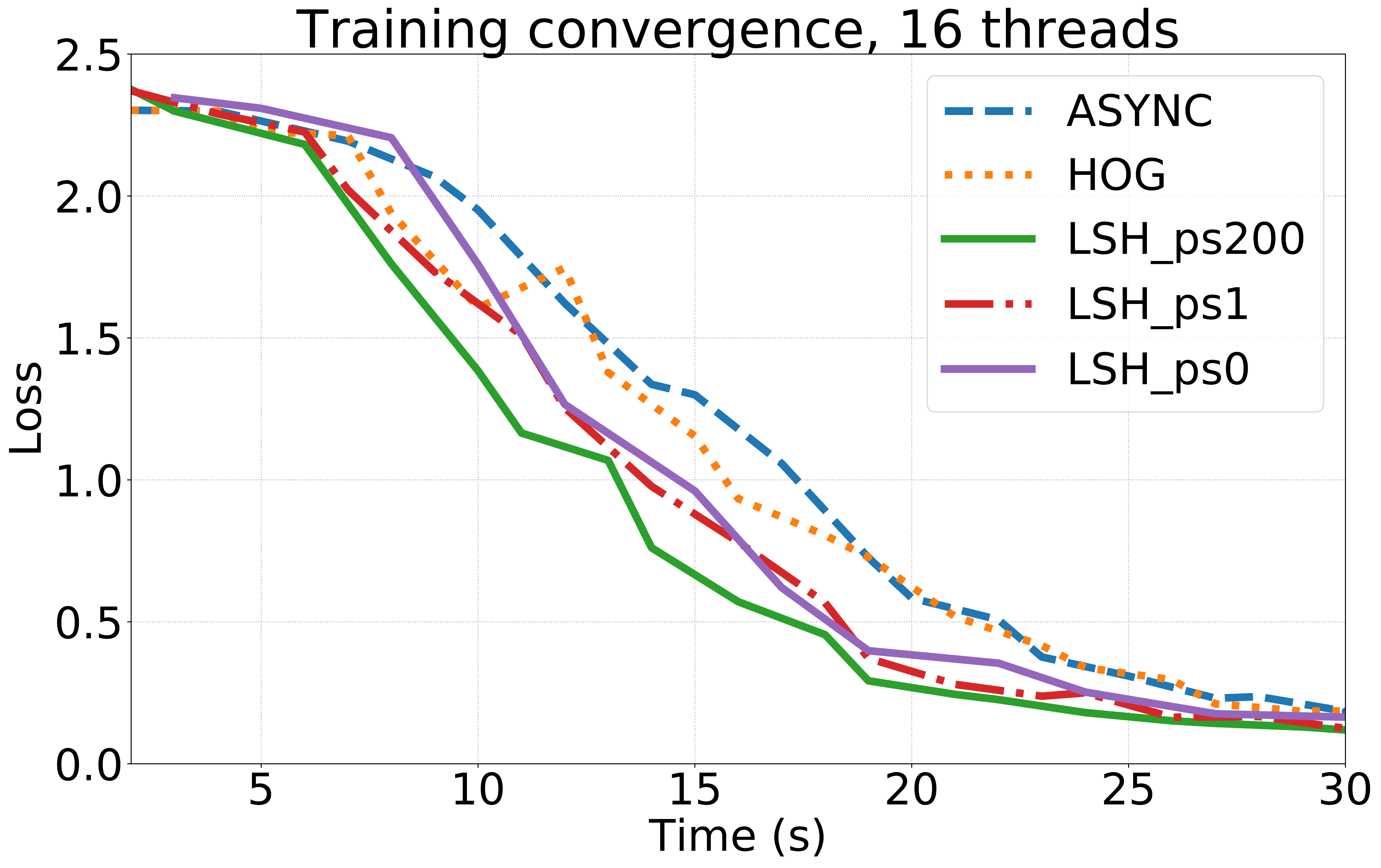}
\end{minipage}
\begin{minipage}[b]{.32\textwidth}
\includegraphics[width=\textwidth]{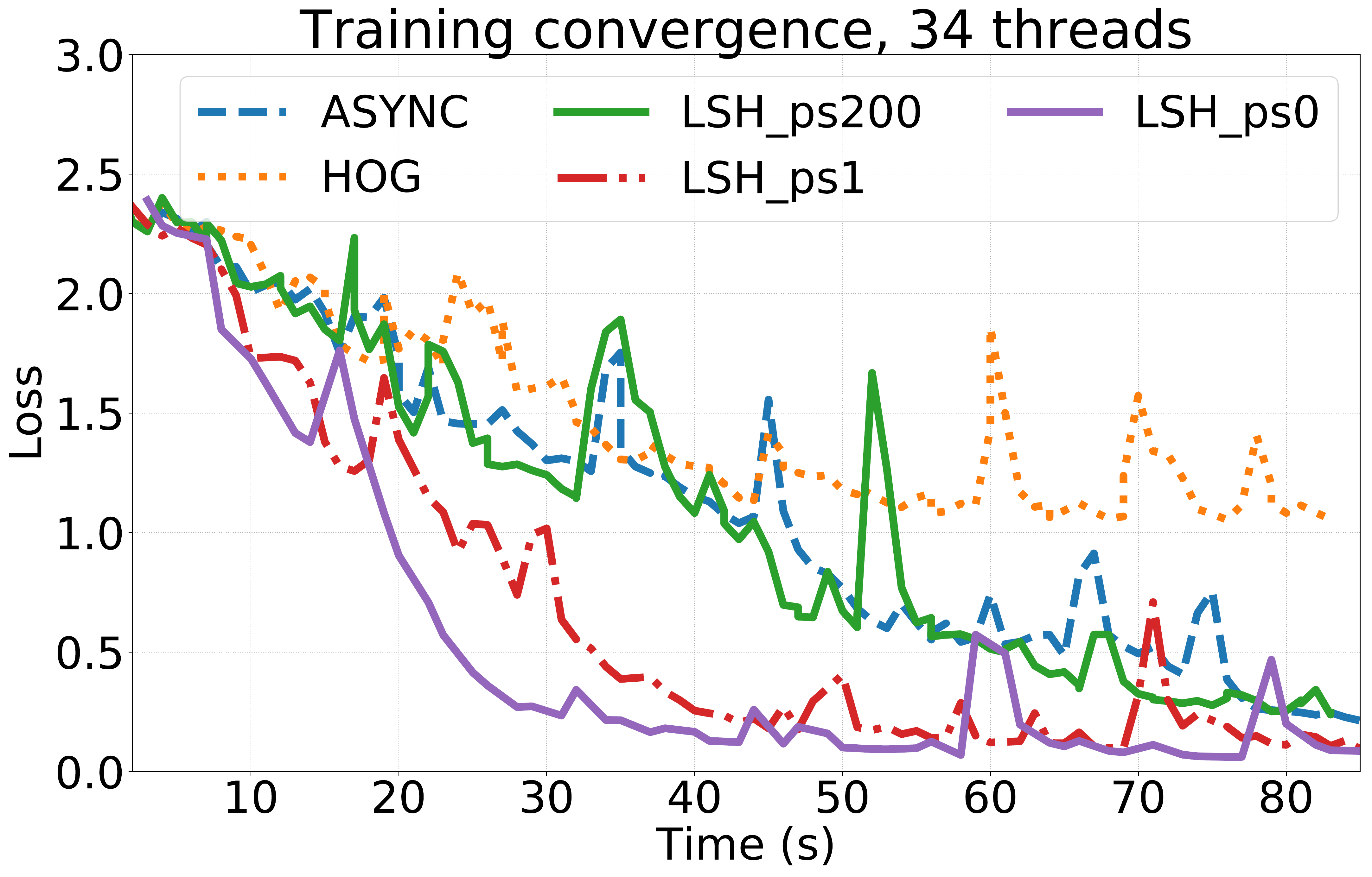}
\end{minipage}
\begin{minipage}[b]{.32\textwidth}
\includegraphics[width=\textwidth]{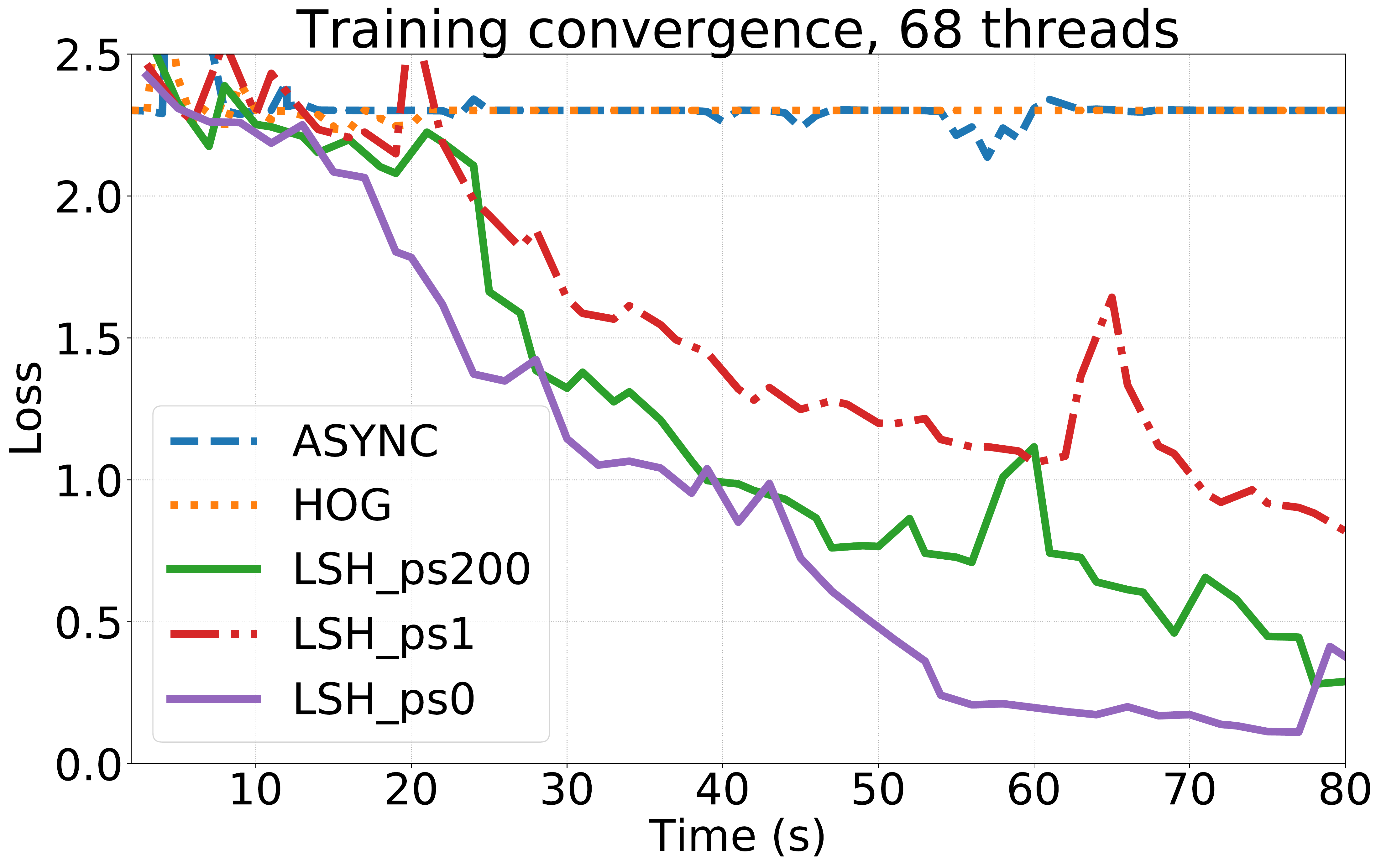}
\end{minipage}
\caption{MLP training progress over time with $m=16$ threads (\textit{left}), with $m=34$ threads (\textit{middle}) and maximum parallelism $m=68$ threads (\textit{right}). The proposed framework ($LSH\_psX$, persistence bound $X$) converges significantly faster relative to baselines (ASYNC, HOG). Under maximum parallelism, the baselines completely fail to converge and oscillate around the initialization point.}\label{fig:trainplot_MLP}
\end{figure*}

\begin{figure*}
\centering
\begin{minipage}[b]{.32\textwidth}
\includegraphics[width=\textwidth]{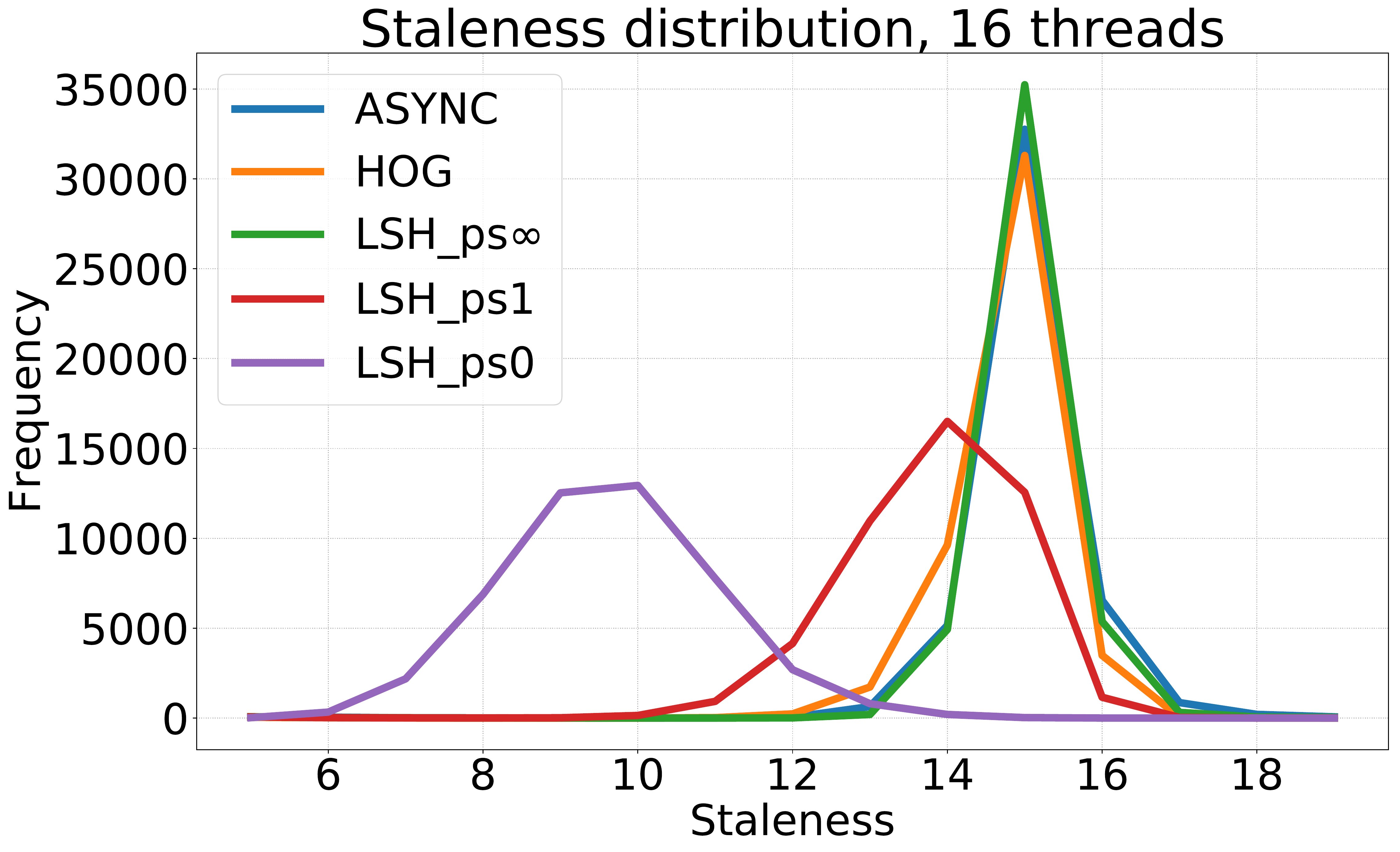}
\end{minipage}
\begin{minipage}[b]{.32\textwidth}
\includegraphics[width=\textwidth]{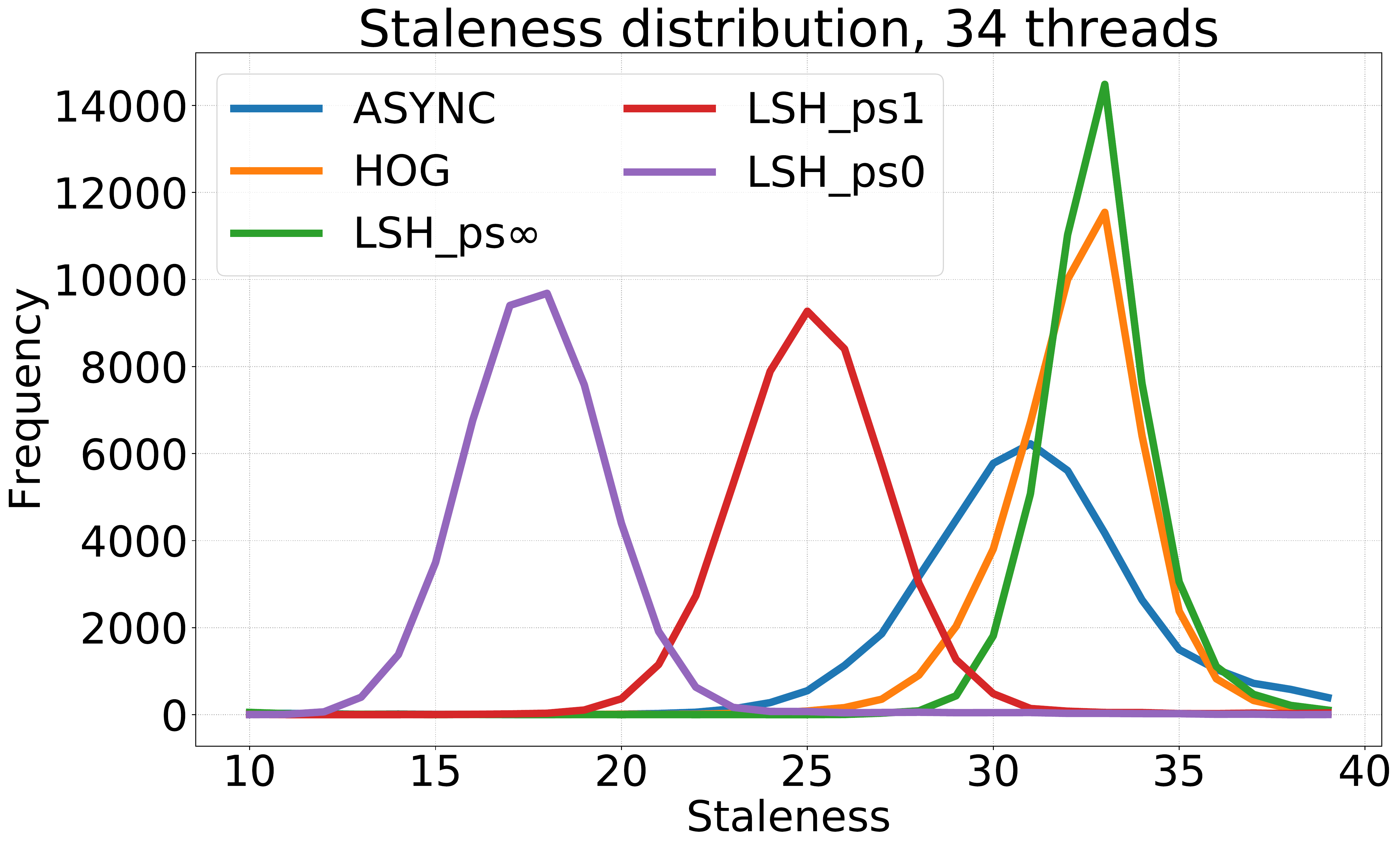}
\end{minipage}
\begin{minipage}[b]{.32\textwidth}
\includegraphics[width=\textwidth]{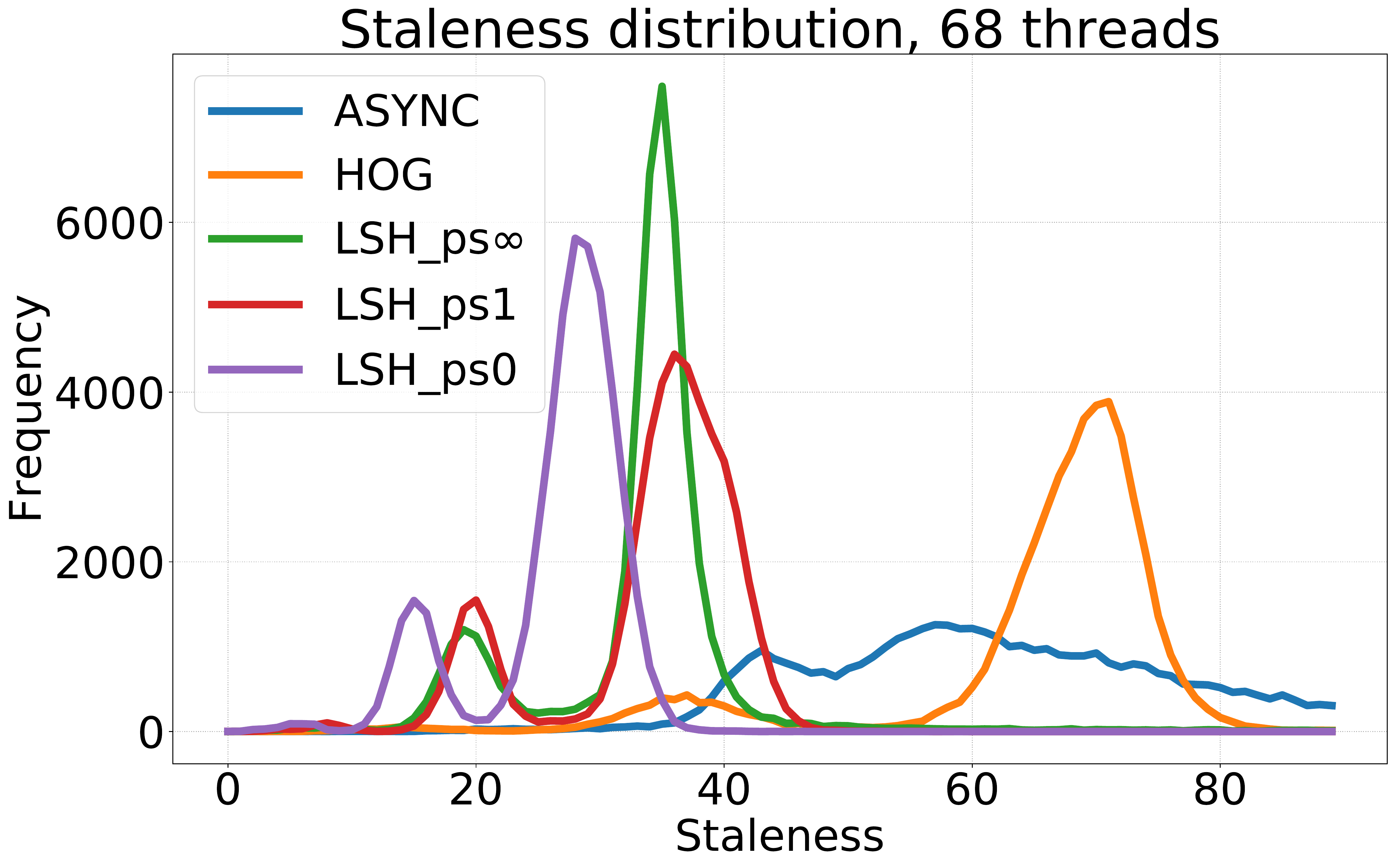}
\end{minipage}
\caption{Staleness distribution over time for MLP with $m=16$ threads (\textit{left}), $m=34$ threads (\textit{middle}) and maximum parallelism $m=68$ threads (\textit{right}). The effect from the contention-regulating persistence bound (\textit{ps}$\in\{0,1,\infty\}$) is clear, and effectively reduces the overall staleness distribution. Under maximum parallelism $m=68$ the ability of the proposed framework (LSH) to self-regulate the balance between latency and throughput becomes clear, with overall lower staleness as well as naturally appearing clusters of threads with higher update rate. The baselines show overall higher staleness distributions, as well as high irregularity for ASYNC due to contention about the locks.}\label{fig:taudist_MLP}
\end{figure*}

\begin{figure*}
\centering
\begin{minipage}[b]{.32\textwidth}
\includegraphics[width=\textwidth]{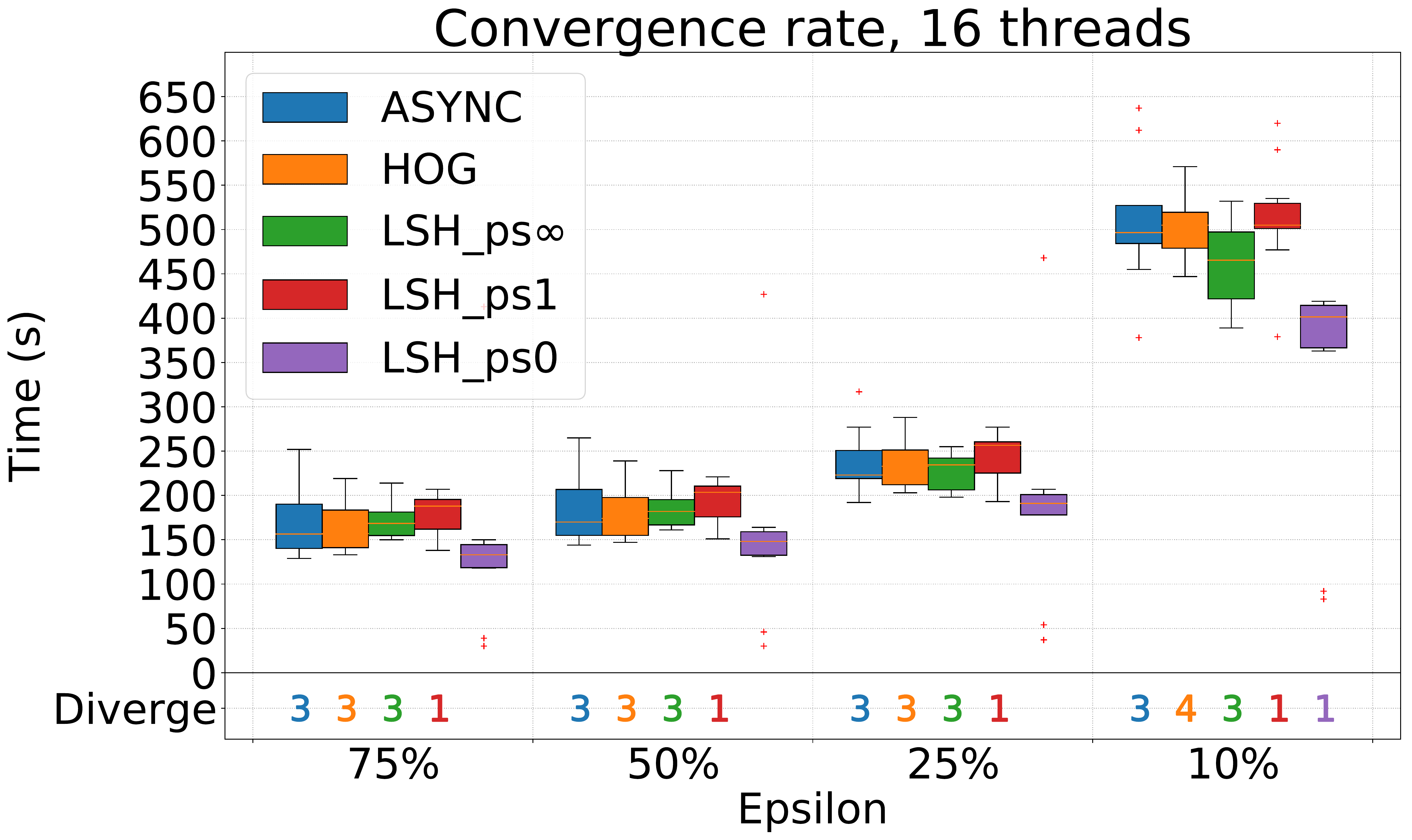}
\end{minipage}
\begin{minipage}[b]{.32\textwidth}
\includegraphics[width=\textwidth]{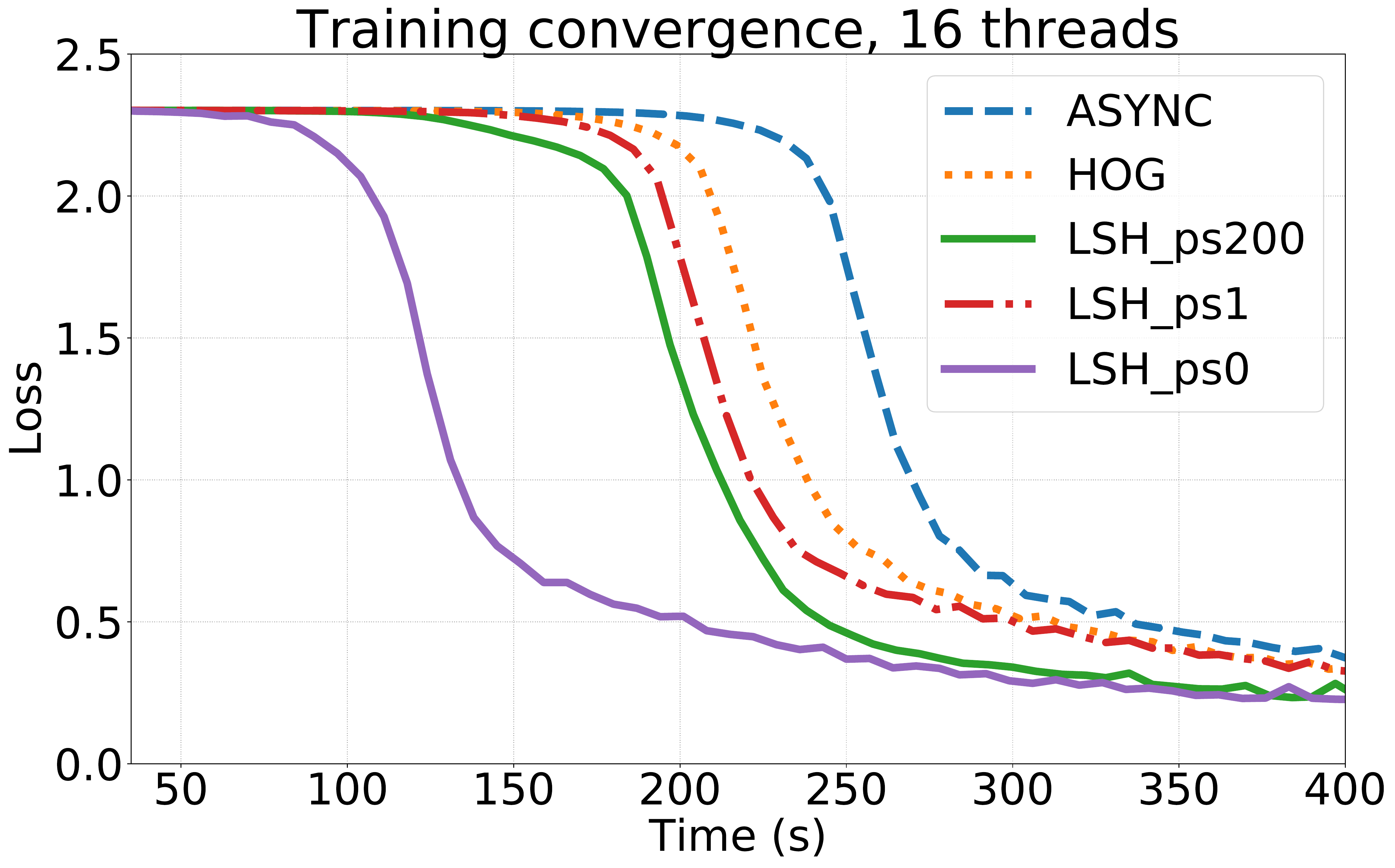}
\end{minipage}
\begin{minipage}[b]{.32\textwidth}
\includegraphics[width=\textwidth]{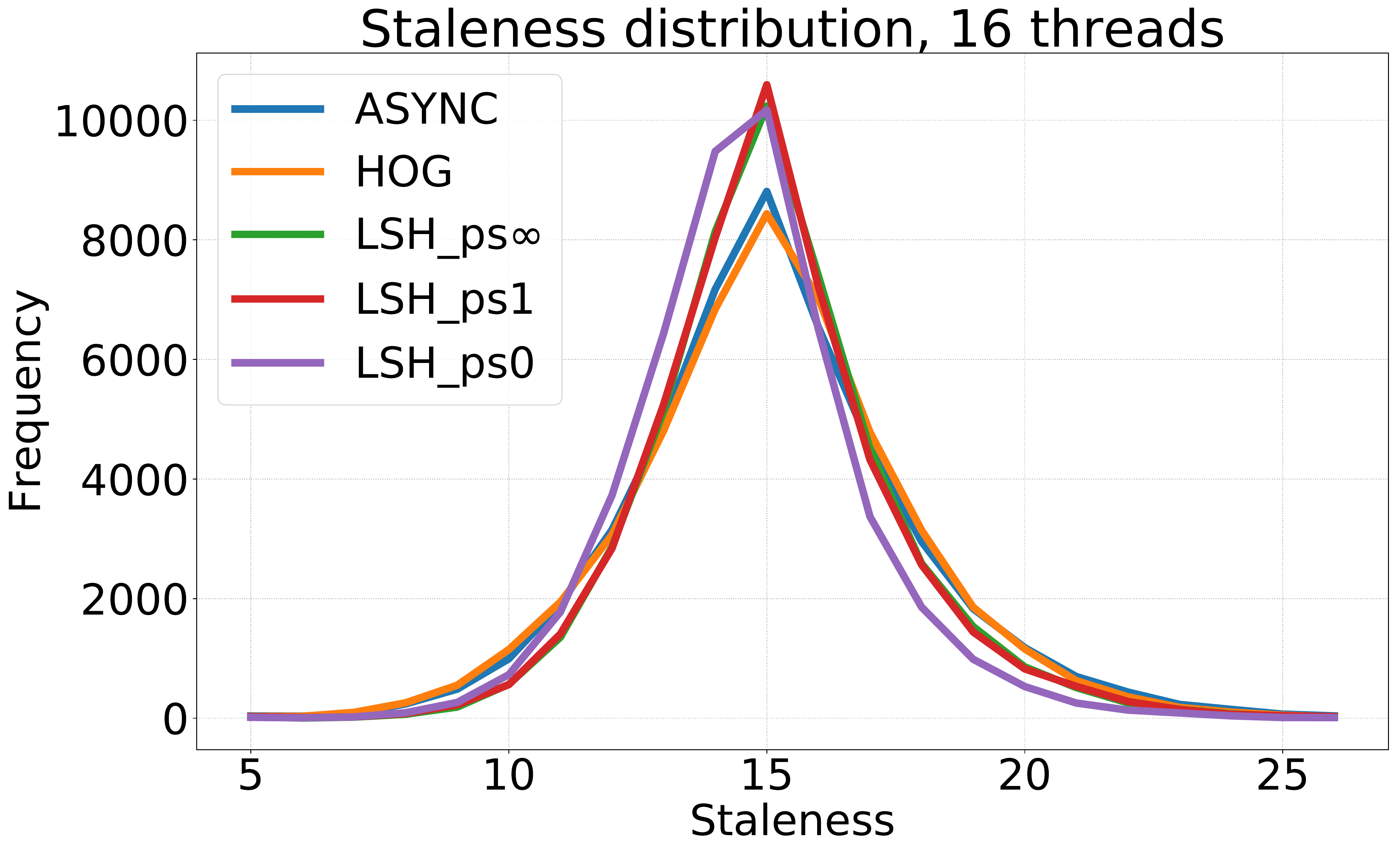}
\end{minipage}
\caption{$\epsilon$-convergence rates to different precision for CNN training with $m=16$ threads; LSH\_ps0 shows $400$s median $10\%$-convergence time, compared to $\sim 500$s for the baselines, with two executions showing remarkable $10\%$-convergence time of below $100$s, i.e. a $4\times$ speedup relative to the best baseline convergence rate of $375$s (\textit{left}) training progress over time (\textit{middle}) and staleness distribution (\textit{right}). The proposed framework (LSH) consistently shows improved convergence rate, as well as solution of lower error.}\label{fig:cnn_tests}
\end{figure*}

\subsubsection{Implementation}
The algorithms and the framework are implemented with C++, with OpenMP \cite{dagum1998openmp} for shared-memory parallel computations, and Eigen \cite{eigenweb} for numerical. 
The framework extends the MiniDNN \cite{yixuan} C++ library for DL. For implementing the \paramcont{} and \algname{}, a substantial refactoring was accomplished, extracting all learnable parameters into a collective data structure, the \paramcont{}. This abstraction forms an interface between SGD algorithm constructions and DL operations, enabling implementation of consistency of different degrees through various synchronization methods. The proposed framework \algname{} application-specific and apply as parallelization of SGD for any optimization problem, in particular of high dimension. For the empirical evaluation an extensible framework is implemented in conjunction with ANN operations, facilitating further research exploring algorithms for parallel SGD for DL with various synchronization mechanisms.\looseness=-1

\subsubsection{Experiment setup}

We evaluate the methods of Section \ref{sec:method} for two DL applications, namely MLP and CNN training on the MNIST benchmarking dataset~\cite{lecun-mnisthandwrittendigit-2010}. The proposed method, however, facilitates generic implementations of SGD, and is applicable over a broad spectrum of optimization problems. We choose to focus the evaluation around benchmarking on DL problems in order to evaluate on relevant applications, as well as to challenge the proposed method, keeping in mind the non-convex and highly irregular nature of the target functions such problems constitute. Moreover, it is in this domain where better understanding of how to support the processing infrastructure is the most needed.
MNIST contains 60,000 images of hand-written digits $\in\{0,\dots9\}$, each belonging to one of ten classes, sampled in mini-batches of 512. Due to space limitations, the details of the MLP and CNN architectures are available in the \suppl{}. The size of the parameter vector $\theta$ are $d=134,794$ and $d=27,354$ for MLP and CNN, respectively.
The experiments are conducted on a 2.10 GHz Intel(R) Xeon(R) E5-2695 system with 36 cores on two sockets (18 cores per socket, each supporting two hyper-threads), 64GB memory, running Ubuntu 16.04.

Box plots in the figures contain statistics ($1^{st}$ and $3^{rd}$ quantiles, minimum and maximum) from 11 independent executions of each setting;
outliers are indicated with the symbol~$+$. 
Where executions fail to reach the required precision $\epsilon$,
the measurement is not included as basis for the box. 
Such execution instances, and those that fail due to numerical instability from staleness, are indicated as 'Diverge' and 'Crash', respectively.
This information is highlighted because failing DL training executions due to noise from staleness or hyper-parameter choices is a common problem in practice \cite{zhang2019empirical}. It is vital that training succeeds, and that the execution time thereby is not wasted.
The threshold $\epsilon$ is specified in terms of percentage of the target function at initialization $f(\theta_0) \approx 2.3$.\looseness=-1

\subsubsection{Experiment outcomes}

The steps of our experiment methodology, summarized in Table \ref{tab:summary_exp}, are as follows: \\
\emph{{\bf S1.} Convergence and hyper-parameter selection:} We benchmark the convergence of the algorithms considered under a wide spectrum of parallelism, and for varying step size $\eta$.
In this step the executions are halted at $\epsilon=50\%$ in order to acquire an overview of the general scalability and relative performance among the evaluated methods.
The results are presented in Fig. \ref{fig:convrate_compeff}, showing a complete picture of the convergence rate and computational efficiency under varying parallelism, the metric of interest being the wall-clock time required until reaching $\epsilon$-convergence.
The baselines are at best with $m=16$ threads and $\eta=0.005$, which we choose as a yardstick for further tests to ensure a fair comparison, and to stress-test \algname{}. The results of the step size test appears in the \suppl{}, showing higher capability of the proposed \algname{} to converge for larger~$\eta$.\looseness=-1 \\
\emph{{\bf S2.} High-precision convergence for MLP:} Using the setting selected according to the above, we benchmark the algorithms and their convergence rate for reaching high precision ($\epsilon=2.5\%)$. We pay attention to the staleness $\tau$ distribution, to gain understanding based also on the results of section~\ref{sec:analysis}.\\
Using $m=16$, $\eta=0.005$, we benchmark \algname{} and baselines to high-precision $2.5\%$-convergence, measuring the wall-clock time (Fig. \ref{fig:precision_MLP}, left). \algname{} shows competitive performance, with faster convergence and smaller fluctuations. In particular, LSH\_ps$\infty$ reaches $\epsilon=2.5\%$ error within $65$s median (compared to baselines' $89$s and $80$s). As hypothesised in section~\ref{sec:analysis}, Fig. \ref{fig:taudist_MLP} confirms that the staleness distribution is significantly reduced by the persistence bound.
\emph{{\bf S3.} Convergence rates for CNN:}
We study the convergence for the CNN application, benchmarking time to convergence for increasing precision $\epsilon$, studying the staleness and convergence over time.
The proposed \algname{} shows fewer diverging executions, with significant improvements in time to high precision convergence with up to $4\times$ speedup relative to the baselines \asyncsgd{} (Fig. \ref{fig:cnn_tests}). Measurements of memory consumption and computation times ($T_c, T_u$) appear in the \suppl{}. Due to the sparse nature of the CNN topology, the gradient computation vs. update application time ratio $T_c/T_u$ is high, leading to a significantly reduced memory footprint (with 17\% on average) of \algname{}.
\looseness=-1\\
\emph{{\bf S4.} Higher parallelization for MLP:} We stress-test the methods, with $m=24$, $m=34$ (max. solo-core parallelism) and $m=68$ (max. hyper-threading).
The results appear in Fig. \ref{fig:precision_MLP}-\ref{fig:taudist_MLP}, showing \algname{} provides significantly improved convergence and stability, with improved staleness. \\
\emph{{\bf S5.} Memory consumption:} We perform a fine-grained continuous measurement of the memory consumption of all algorithms considered, for MLP and CNN training.  For the CNN application, \algname{} reduces the memory consumption by $17\%$ on average thanks to dynamic allocation of \paramcont{} and efficient memory recycling.The detailed plots appear in the \suppl{}.

\subsubsection{Summary of outcomes}
\algname{} shows overall an improved convergence rate, stable under varying parallelism and hyper-parameters, and significantly fewer executions that fail to achieve $\epsilon$-convergence.
In presence of contention, the lock-free nature enables \algname{} to self-regulate the balance between throughput and latency, and converge in settings where the baselines fail completely. Even the case that with $T_p=\infty$, i.e. without starvation-freedom, we see persistent improvements relative to the baselines, demonstrating in this demanding context too, a useful property, namely that lock-freedom balances between system-wide throughput and thread-associated latency ~\cite{cederman2013study,gulisano2017efficient}.\looseness=-1

\section{Related Work} \label{sec:related_works}

The study of numerical methods under parallelism sparked due to the works by Bertsekas and Tsitsiklis~\cite{bertsekas1989parallel}.
Distributed and parallel asynchronous SGD has since been an attractive target of study, e.g.~\cite{duchi2015asynchronous,sallinen2016high,chaturapruek2015asynchronous,lian2015asynchronous}, among which \hogwild{}~\cite{recht2011hogwild}.
In the recent \cite{alistarh2020elastic} the concept of bounded divergence between the parameter vector and the threads' view of it is introduced, proving convergence bounds for convex and non-convex problems.
De Sa et. al \cite{de2015taming} introduced a framework for analysis of \hogwild{}-style algorithms. This was extended in \cite{alistarh2018podc}, showing the bound increases with a magnitude of $\sqrt{d}$ due to inconsistency, implying higher statistical penalty for high-dimensional problems. This strongly motivates studying algorithms which, while enjoying the computational benefits of lock-freedom, also ensure consistency. To our knowledge, this has not been done prior to the present work.\looseness=-1

In \cite{mania2017perturbed} the algorithmic effect of asynchrony in \asyncsgd{} is modelled by perturbing the stochastic iterates with bounded noise. Their framework yields convergence bounds, but as described in the paper, are not tight, and rely on strong convexity.\looseness=-1

In \cite{ma2019stochastic}, with motivation related to ours, a detailed study of parallel SGD focusing on \hogwild{} and a new, GPU-implementation, is conducted, focusing on convex functions, with dense and sparse data sets and comparison of different computing architectures.
Here  we propose an extensible framework
of consistency-preserving algorithmic implementations
of \asyncsgd{} together with \hogwild{}, 
that covers 
the associated design space of \asyncsgd{} algorithms, and we focus on  MLP and CNN, 
which are inherently more difficult to parallelise.\looseness=-1

In 	\cite{wei2019automating}, as in this work, the focus is the  fundamental limitation of data parallelism in ML. 
They, too,  point out that the limitations are due to
concurrent SGD parameter accesses,
usually diminishing or even negating the parallelisation
benefits. 
To alleviate this, they propose the use of static analysis for identification of data that do not cause dependencies, for parallelising their access. They do this as part of a system that uses Julia, a script language that performs just-in-time compilation. Their approach is effective and works well for e.g. Matrix factorization SGD. For DNNs, that we consider in this paper, as they explain, their work is not directly applicable, since in DNNs permitting "good" dependence violation is the common parallelization approach. \looseness=-1

There are works introducing adaptiveness to staleness \cite{mcmahan2014delay,zhang2016staleness,sra2015adadelay} and in particular in \cite{backstrom2019mindthestep} for a deep learning application. This research direction is orthogonal to this work and can be applied in conjunction with the algorithms and synchronization mechanisms considered here.

Asynchronous SGD approaches for DNNs are scarce in the current literature. In the recent work \cite{lopez2020asynchronous}, Lopez et al. propose a semi-asynchronous SGD variant for DNN training, however requiring a master thread synchronizing the updates through gradient averaging, and relying on atomic updates of the entire parameter vector, resembling more a shared-memory implementation of parameter server.
In \cite{stich2019local} theoretical convergence analysis is presented for \syncsgd{} with \textit{once-in-a-while} synchronization. They mention the analysis can guide in applying \syncsgd{} for DL, however the analysis requires strong convexity.
\cite{jiang2017collaborative} proposes a consensus-based SGD algorithm for distributed DL. They provide theoretical convergence guarantees, also in the non-convex case, however the empirical evaluation is limited to iteration counting as opposed to wall-clock time measurements, with mixed performance positioning relative to the baselines.
In \cite{lian2018asynchronous} a topology for decentralized parallel SGD is proposed, using pair-wise averaging synchronization.
In the recent \cite{li2020taming} a partial all-reduce relaxation of \syncsgd{} is proposed, showing improved convergence rates in practice when synchronizing only subsets of the threads at a time, due to higher throughput, complemented with convergence analysis for convex and non-convex problems. In particular, the empirical evaluation shows only requiring one thread (i.e. \asyncsgd{}) gives competitive performance due to the \textit{wait-freedom} that follows from the lack of synchronization.

\section{Conclusions} \label{sec:conclusion}

We propose the extensible generic algorithmic framework \algname{} for asynchronous lock-free parallel SGD, together with \paramcont{}, a data type providing an abstraction of common operations on high-dimensional model parameters in ANN training, facilitating modular further exploration of aspects of parallelism and consistency, connecting to and extending work in the literature on bulk operations on container data structures~\cite{nikolakopoulos2015consistency}.

We analyze safety and progress guarantees of the proposed \algname{}, as well as bounds on the memory consumption, execution dynamics, and contention regulation.
Aiming at understanding the influence of synchronization methods for consistency of shared data in parallel SGD, we provide a comprehensive empirical study of \algname{} and established baselines, benchmarking on two prominent \textit{deep learning} (DL) applications, namely MLP and CNN for image classification. The benchmarks are chosen in order to challenge the proposed model against  the baselines, and provide new useful insights in the applicability of \asyncsgd{} in practice.
We observe that the baselines, i.e. standard implementations of \asyncsgd{}, are very sensitive to hyper-parameter choices and are prone to unstable executions due to noise from staleness. The proposed framework \algname{} outperforms the baselines where they perform the best, and provides a balanced behaviour, implying stable and timely convergence for a far wider spectrum of parallelism.

The methods are implemented in an extensible C++ framework, interfacing DL operations with parallel SGD algorithms, facilitating further research, exploring algorithms for parallel SGD for DL with various synchronization mechanisms. Exploring different consistency types for the $theta$ updates, in conjunction to sparcification approaches are interesting directions to pursue. \looseness=-1

\begin{footnotesize}
\bibliographystyle{abbrv}
\bibliography{ref}
\end{footnotesize}

\newpage
\appendix

\section*{On MLPs and CNN}
    \textbf{\em MLPs} consist of several stacked densely-connected layers of neurons, each applying a non-linear transformation of the input and passing the result to the next layer:
    \begin{align*}
        o_n^{(l)} &= \sigma \left( \sum_{i=0}^{|N_{l-1}|-1} \theta_i^{(l,n,w)} \cdot o_i^{(l-1)} + \theta^{(l,n,b)} \right)
    \end{align*}
    where $o_n^{(l)}$ is the output of neuron $n \in \{0,\dots,N_l-1\}$ in the $l^{th}$ layer, $\sigma$ is a non-linear activation function, typically the ReLU function $\sigma(x) = max(0,x)$, and $\theta^{(l,n,w)}, \theta^{(l,n,b)}$ contains the learnable weights and bias parameters of to the $n^{th}$ neuron.\\
    %\item 
    \textbf{\em CNNs} consist of \textit{convolutional} layers, convolving the input with learnable filters for feature detection:
    \begin{align*}
        o_{n,f}^{(l)} &= \sigma \left( \sum_{i=0}^k \theta_i^{(l,f,w)} \cdot o_{n+i}^{(l-1)} + \theta^{(l,f,b)} \right)
    \end{align*}
    for a number of filters $f$, corresponding to a 1D convolution, but can be naturally extended to 2D. Convolutional layers are sparsely connected, reducing the number of weights to be trained, and are especially efficient for analysis of image/spatial data due to the translation-invariant property of feature detection with convolution.
Convolutional layers are often used in combination with \textit{MaxPool} layers, which map the output of a number of consecutive neurons onto their maximum. This significantly reduces dimension of the signal and the learnable weights.

We refer to the collection of all parameters $\theta^{(l,n,w/b)}$, $\theta^{(l,f,w/b)}$ belonging to an ANN flattened into a 1D array as the \textit{parameter vector}, denoted as $\theta_t$, at iteration $t$ of SGD. This abstraction is used in subsequent sections when arguing regarding consistency and progress.

In the output layer of an ANN, the \textit{softmax} activation function $\sigma_i(x) = e^{x_i}/\sum^{|x|}_{j=1} e^{x_j}$, for each output neuron $i$, is often used for classification problems, outputting an estimated class distribution $y$ of an input $x$. Given the true class/label $\hat{y}$, the ANN performance is quantified by the cross-entropy loss function:\looseness=-1
\begin{align*}
    L(\hat{y},y(x:\theta)) = - \sum_i^{|out|} y(x:\theta)_i \log(\hat{y}_i) \label{eq:crossentropy}
\end{align*}
where $y$ contains the outputs from the last layer, and depends on the input $x$ and the current state of $\theta$. The \textit{training} process for ANNs then constitutes of iteratively adjusting $\theta$ to minimize the error function $f(\theta) = L(\hat{y},y(x:\theta))$. The \textit{BackProp} algorithm is used for computing $\nabla_\theta f(\theta)$, and SGD is then used for minimizing $f$, and training the ANN. In every iteration the input is selected at random, either as single data point or as a \emph{batch} considered in conjunction.

\section*{Analysis - complementary material}

\setcounter{algocf}{3}

\begin{algorithm}
    \scriptsize

    GLOBAL ParamVector $PARAM$ \hfill \\
    GLOBAL Float $\eta$ \hfill // step size \\
    
    \vspace{1pt}
    
    \underline{Initialization}\;
    
    $PARAM \leftarrow $ new $ParamVector()$ \\
    $PARAM.\randinit{}$ \hfill // randomly initialize parameters \\
    
    \vspace{1pt}
    
    \underline{Each thread}\;
    
    $local\_grad \leftarrow $ new $ParamVector()$ \hfill // local gradient memory \\
    $local\_param \leftarrow $ new $ParamVector()$ 
    
    \Repeat {convergence}{
        $local\_param.theta = $ copy$(PARAM.theta)$ \\
        $local\_grad.theta \leftarrow comp\_grad(local\_param.theta)$  \hfill // gradient\\
        $PARAM$.\PCupdate{$local\_grad.theta, \eta$} \\
    }
    
    \caption{\hogwild{} expressed using the \paramcont{} interface}
    \label{algorithm:hogwild_implementation}
    
    \vspace{2pt}
\end{algorithm}

\textit{Proof sketch - Lemma 2:}
The first claim (i) follows from the definition of the safe\_delete operation of the \paramcont{}, ensuring that the memory of an instance $PC_t$ is reclaimed only if $stale\_flag=true$ ($P$ points to a newer instance, ensuring no new readers of $PC_t$), $n\_readers=0$ (no readers currently) and that the memory has not already been reclaimed.
The second claim (ii) is realized by the fact that the memory recycling mechanism is exhaustive, i.e. \paramcont{} instances that will not be used further by any thread will eventually be reclaimed through the delete operation in line 10 of Algorithm 1. The reason is the following: each thread that finishes its use of a \paramcont{} instance will call the  stop\_reading operation, which in turn calls safe\_delete, which reclaims the memory if safe, according to the above, i.e. it holds that the instance is currently not in use and will not be in the future. If that is not the case, then the threads that are currently using the instance will each eventually invoke the safe\_delete operation, the last of which will perform the reclamation. Now, from Algorithm 3 it is clear that in the worst case each thread has a unique $latest\_param$ on which it is active reader, and an additional two \paramcont{} ($new\_param$ and $local\_grad$), giving in total $3m$. $\hfill\blacksquare$

%\section*{Supplementary Material on Contention and Staleness}

\textit{Proof of Theorem 3}
From (4), we have
\begin{align*}
    n_{t} &= n_{t-1} + \frac{m-n_{t-1}}{T_c} - \frac{n_{t-1}}{T_u} \\
    &= (1 - 1/T_c - 1/T_u) n_{t-1} + m/T_c \\
    &= \dots \\
    &= \frac{m}{T_c}\sum_{i=0}^{t-1}(1-1/T_c-1/T_u)^i + (1-1/T_c-1/T_u)^t n_0 \\
    &= \frac{m}{T_c} \frac{1 - (t-1/T_c-1/T_u)^t}{1/T_c+1/T_u} + (1-1/T_c-1/T_u)^t n_0
\end{align*}
$\hfill\blacksquare$

\section{Evaluation - complementary material}

The details of the ANN architectures implemented in the evaluation (Section 5) are shown in Table \ref{tab:MLP_architecture} and \ref{tab:CNN_architecture} for MLP and CNN, respectively.

\begin{table}[htbp]
\small
\begin{center}
\begin{tabular}{|c|c|c|c|}
\hline
\textbf{Layer}&\multicolumn{3}{|c|}{\textbf{Layer details}} \\
\cline{2-4} 
\textbf{\#} & \textbf{\textit{Type}} & \textbf{\textit{\# Neurons}} & \textbf{\textit{Act. fcn.}} \\
\hline
1-3 & Dense & 128 & ReLU \\
4 & Dense & 10 & Softmax \\
\hline
\end{tabular}
\end{center}
\caption{MLP Architecture, $d=134,794$}
\label{tab:MLP_architecture}
\end{table}

\begin{table}[htbp]
\small
\begin{center}
\begin{tabular}{|c|c|c|c|c|c|}
\hline
\textbf{Layer}&\multicolumn{5}{|c|}{\textbf{Layer details}} \\
\cline{2-6} 
\textbf{\#} & \textbf{\textit{Type}} & \textbf{\textit{\# Filters}} & \textbf{\textit{\# Neurons}} & \textbf{\textit{Kernel}} & \textbf{\textit{Act. fcn.}} \\
\hline
1 & Conv$^{\mathrm{a}}$ & 4 & - & (3,3) & ReLU \\
2 & Pool$^{\mathrm{b}}$ & - & - & (2,2) & ReLU \\
3 & Conv$^{\mathrm{a}}$ & 8 & - & (3,3) & ReLU \\
4 & Pool$^{\mathrm{b}}$ & - & - & (2,2) & ReLU \\
5 & Dense & - & 128 & - & ReLU \\
6 & Dense & - & 10 & - & Softmax \\
\hline
\multicolumn{6}{l}{$^{\mathrm{a}}$\textit{Convolutional layer} \ $^{\mathrm{b}}$\textit{MaxPool layer}} \\
\end{tabular}
\end{center}
\caption{CNN Architecture, $d=27,354$}
\label{tab:CNN_architecture}
\end{table}

\subsubsection*{Convergence and hyper-parameter selection}
\begin{figure*}
\centering
\begin{minipage}[b]{.48\textwidth}
\includegraphics[width=\textwidth]{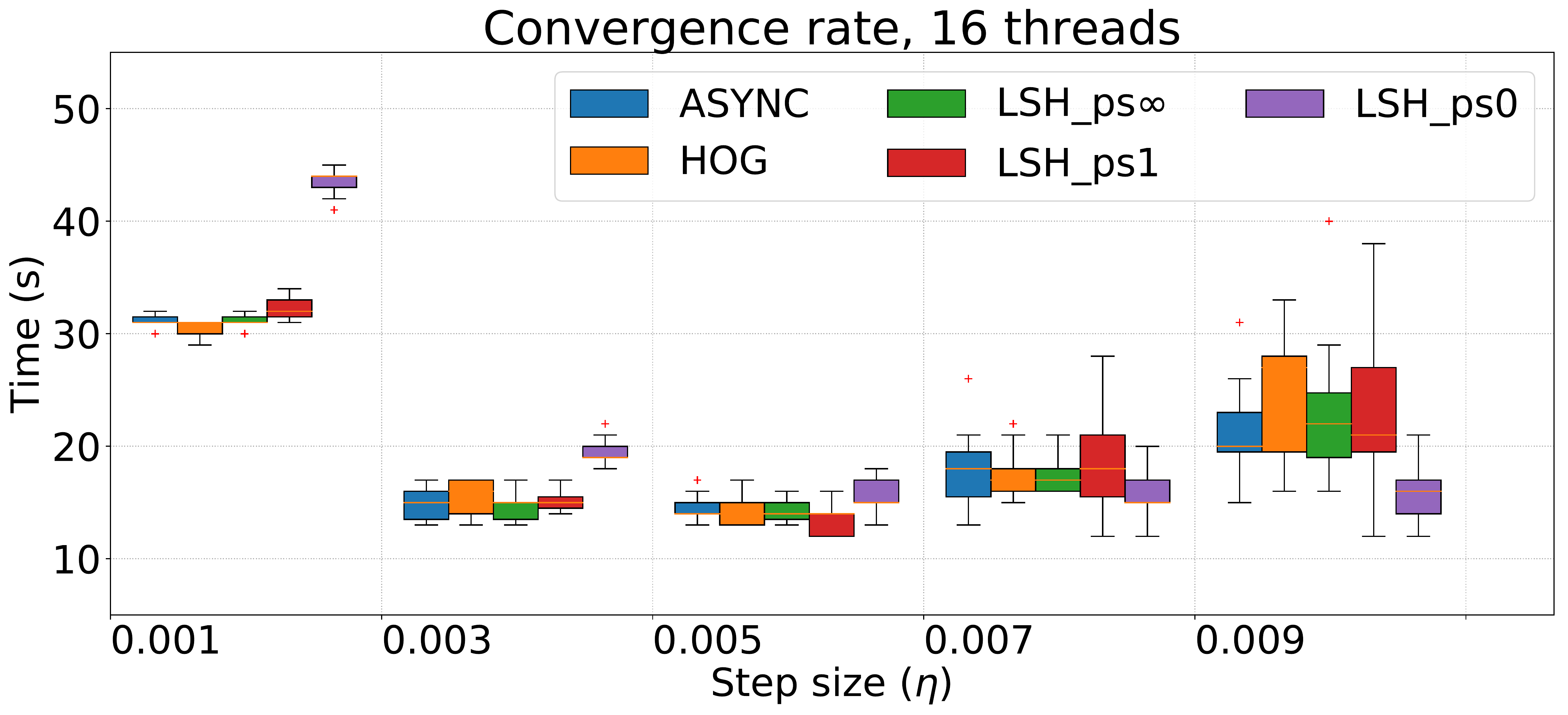}
\end{minipage}\qquad
\begin{minipage}[b]{.48\textwidth}
\includegraphics[width=\textwidth]{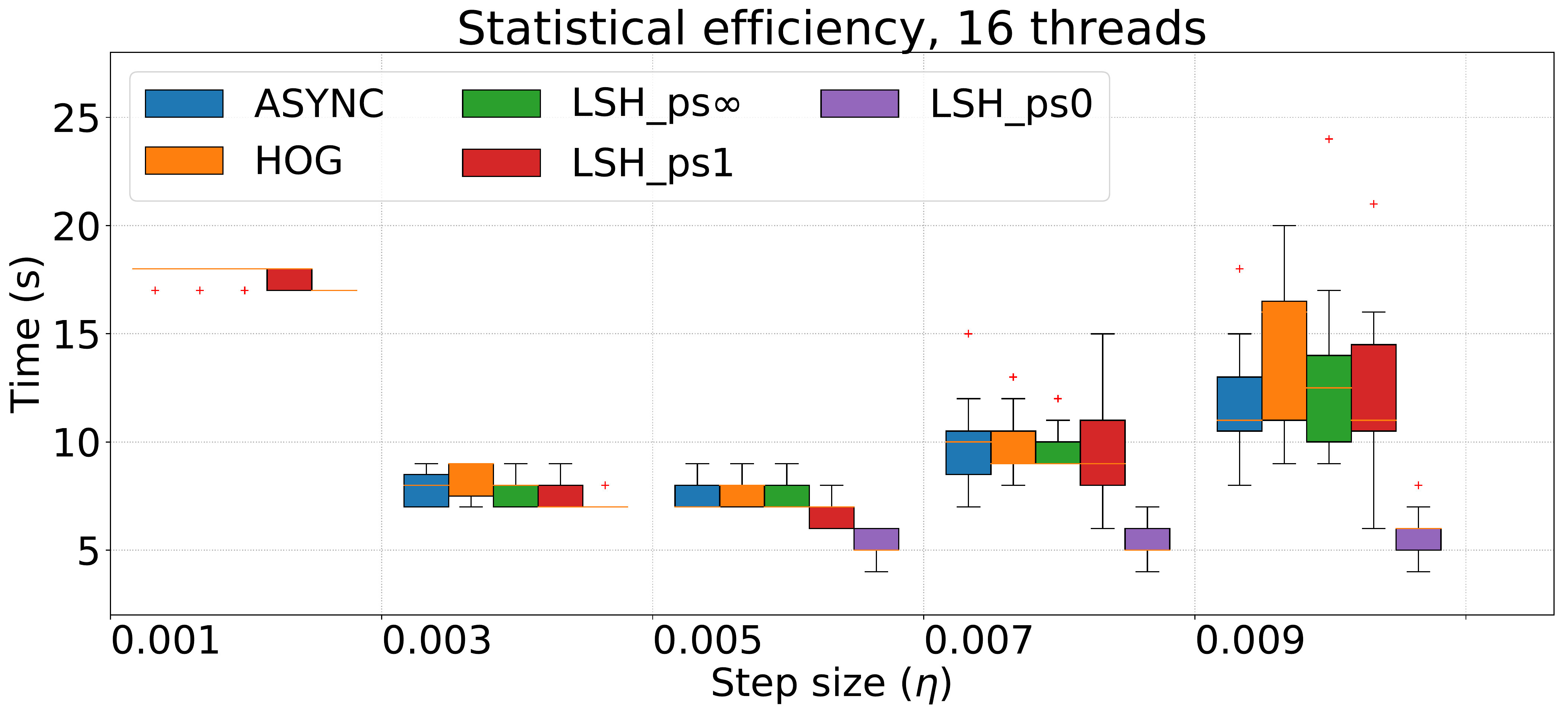}
\end{minipage}
\caption{Step size tuning (left), confirming the choice $\eta=0.005$, and statistical efficiency (right), showing $50\%$-convergence}\label{fig:stepsize_tuning}
\end{figure*}

Figure \ref{fig:stepsize_tuning} shows the convergence rate for different values of step size $\eta$. The baselines \asyncsgd{} and \hogwild{} show the best performance for $\eta=0.005$, which is hence used in the subsequent test stages.

\subsubsection*{Gradient computation and update time - $T_c, T_u$}
\begin{figure*}
\centering
\begin{minipage}[b]{.48\textwidth}
\includegraphics[width=\textwidth]{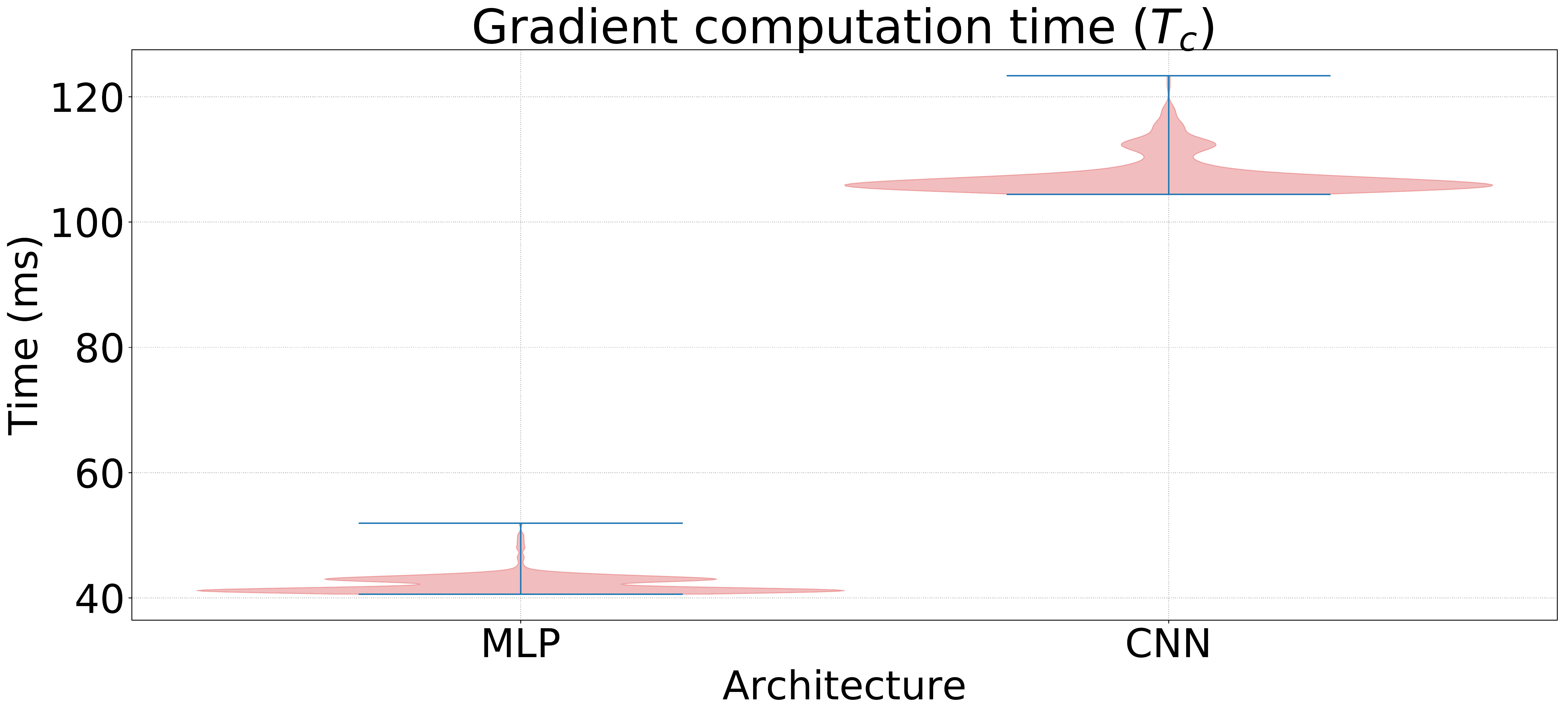}
\end{minipage}\qquad
\begin{minipage}[b]{.48\textwidth}
\includegraphics[width=\textwidth]{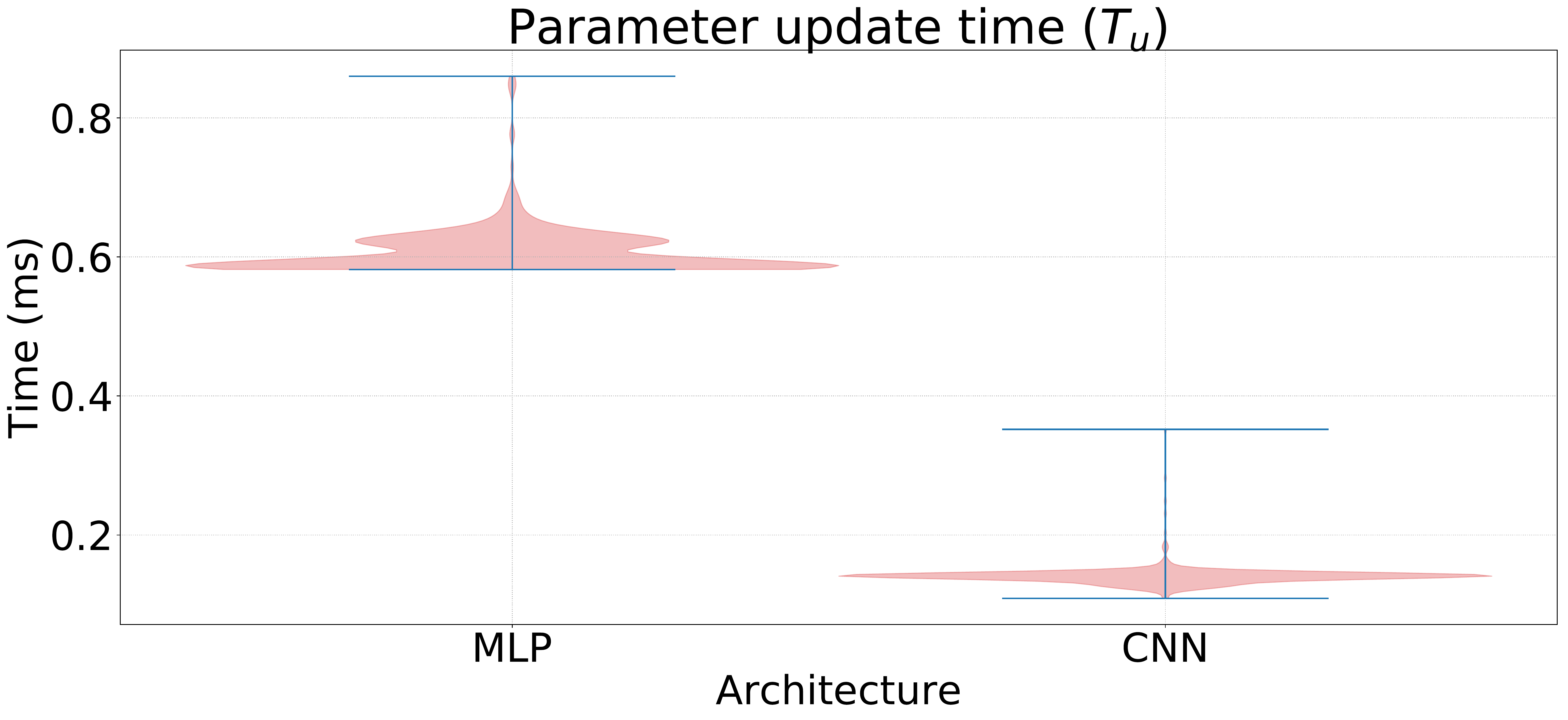}
\end{minipage}
\caption{Gradient computation and parameter update times $T_c, T_u$ (left, right, respectively) for MLP and CNN}\label{fig:Tc_Tu}
\end{figure*}

The distribution of the wall-clock time to compute and apply gradients, respectively, are shown in Figure~\ref{fig:Tc_Tu}. Despite having a lower dimensionality, the gadient computation time $T_c$ is higher for CNN. This is due to the topological nature of the convolutional layer, where filters are strided along the input image pixel by pixel. This requires in practice a large number of smaller matrix multiplications, as opposed to MLP which instead consists of few but significantly larger ones. However, the time to apply one gradient $T_u$ is smaller in the CNN application, since the $\theta$ vector is smaller.

Since the dimension $d$ of the \paramcont{} is significantly smaller for the CNN ($d=27,354$) compared to the MLP ($d=134,794$), the time $T_u$ to apply an update is smaller, but due to the topological nature CNNs, the gradient computation time $T_c$ is relatively high. This results in lower contention in the \casretryloop{}. As a consequence the contention-regulating effect of the \algname{} algorithms does not kick in, hence showing similar staleness distribution as the baselines. The proposed \algname{} nevertheless shows significant improvement in the convergence rate.

\begin{figure*}
\centering
\begin{minipage}[b]{.48\textwidth}
\includegraphics[width=\textwidth]{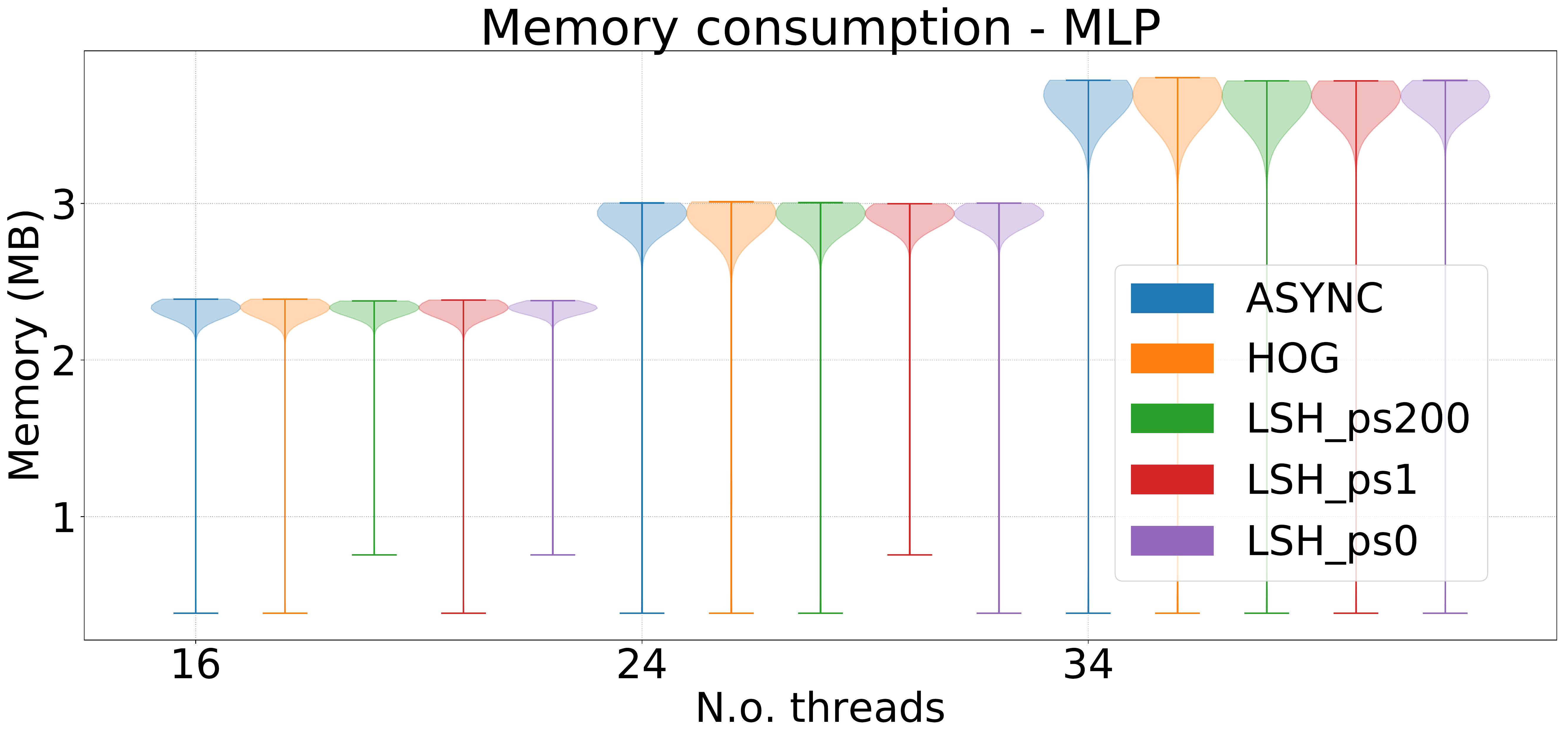}
\end{minipage}\qquad
\begin{minipage}[b]{.48\textwidth}
\includegraphics[width=\textwidth]{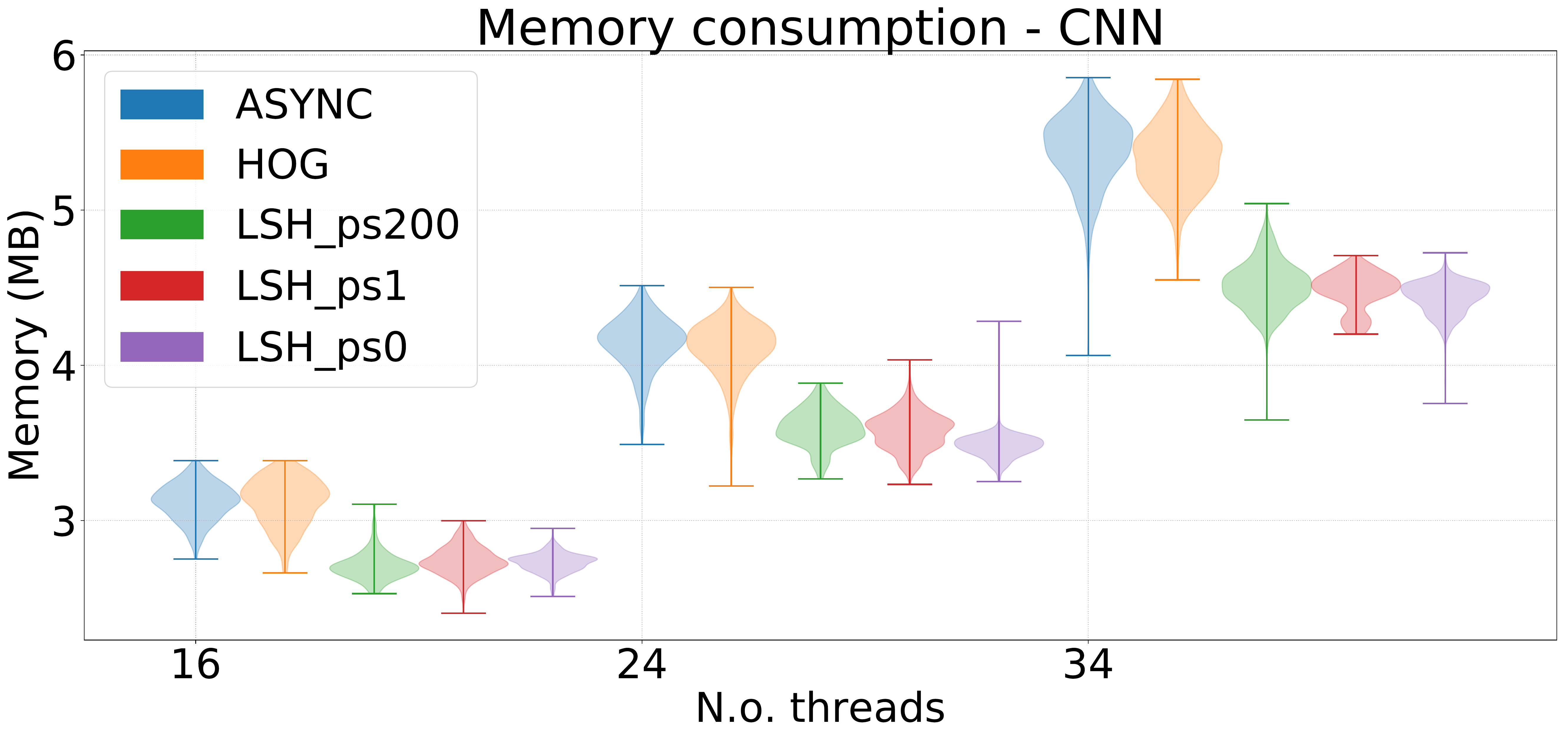}
\end{minipage}
\caption{Memory consumption measured continuously on second granularity for MLP (left) and CNN (right)}\label{fig:mem_con}
\end{figure*}

\subsubsection*{Memory consumption}

Figure \ref{fig:mem_con} shows the distribution of the memory consumption of the different algorithms for MLP and CNN training. The measurements were acquired using the UNIX \texttt{ps} command, collected with second granularity.

\end{document}